\documentclass[lettersize,journal]{IEEEtran}

\usepackage{amsmath,amsfonts}
\usepackage{array}
\usepackage[caption=false,font=normalsize,labelfont=sf,textfont=sf]{subfig}
\usepackage{textcomp}
\usepackage{url}
\usepackage{graphicx}
\usepackage{cite}
\usepackage{booktabs}
\usepackage[table,xcdraw]{xcolor}
\usepackage{dblfloatfix}

\usepackage{hyperref}
\hypersetup{
	colorlinks=true,
	linkcolor=blue,
	citecolor=blue,
	urlcolor=blue
}

\makeatletter
\@removefromreset{table}{section}
\makeatother
\begin{document}

\title{%
	Chemical and geometric representation fidelity improves drug--target affinity prediction
}

\author{
	Yixiao Li, Yining Qian, Yefan Chen, Zenghui Chen, Jiayue Sun,
	Yuhai Zhao, Cheng Tan, and An-Yang Lu%
	\thanks{Yixiao Li, Yefan Chen, Zenghui Chen, Jiayue Sun, and An-Yang Lu are with the College of Information Science and Engineering, Northeastern University, Shenyang 110819, China.}%
	\thanks{Yining Qian is with the School of Computer and Communication Engineering, Northeastern University at Qinhuangdao, Qinhuangdao 066004, China.}%
	\thanks{Yuhai Zhao is with the School of Computer Science and Engineering, Northeastern University, Shenyang 110819, China.}%
	\thanks{Cheng Tan is with Shanghai Artificial Intelligence Laboratory, Shanghai 200232, China.}%
	\thanks{Corresponding authors: qianyiningning@126.com (Y.Q.); luanyang@mail.neu.edu.cn (A.-Y.L.).}%
}

\maketitle

\begin{abstract}
Predicting drug--target binding affinity (DTA) requires models to distinguish subtle chemical and structural determinants underlying molecular recognition. Although recent approaches increasingly incorporate richer drug and protein information, such information may be compressed, homogenized or discretized during representation construction, causing affinity-relevant distinctions to be lost before interaction modelling. We hypothesized that this representation-stage information loss constitutes an upstream bottleneck that cannot be reliably overcome by increasingly complex interaction predictors. To test this hypothesis, we developed ReGeoDTA, a representation-preserving framework that maintains affinity-relevant chemical heterogeneity in molecular representations and continuous geometric relationships in protein structures. Across three benchmark datasets, ReGeoDTA consistently improved affinity prediction, and the proposed representation-preserving strategies retained their benefits across diverse DTA architectures. Controlled representation degradation progressively reduced predictive performance, whereas increasing downstream predictor complexity failed to recover information lost during representation construction. These findings identify representation fidelity as an upstream design principle for accurate and generalizable drug--target affinity prediction, with potential implications for computational compound prioritization.
\end{abstract}


\noindent\textbf{TOC Text:}
Representation-stage information loss is identified as a general bottleneck in drug--target affinity prediction. By preserving chemical heterogeneity in drug representations and continuous geometry in protein structures, ReGeoDTA demonstrates that improved representation fidelity consistently benefits diverse DTA architectures, supporting information preservation as a general design principle rather than an architecture-specific route to performance improvement.

\section{Introduction}
	
\begin{figure*}[t]
	\centering
	\includegraphics[width=\textwidth]{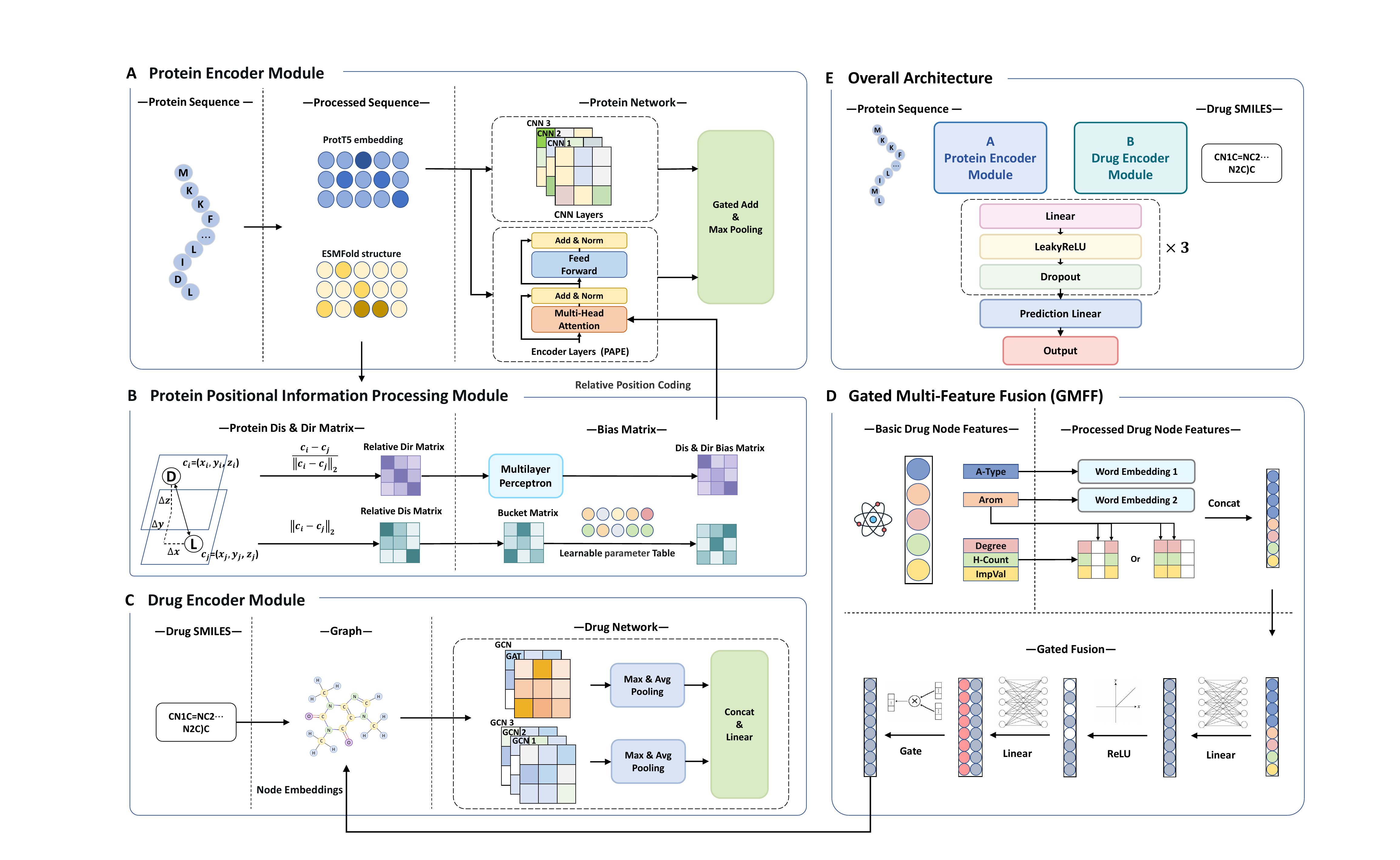}
	\caption{\textbf{Overview of ReGeoDTA. A} Position-Aware Protein Encoder (PAPE) with relative geometric positional encoding. \textbf{B} Protein positional information processing module for generating processed relative geometric information from 3D protein coordinates. \textbf{C} Drug graph encoder with GMFF-based node feature representation. \textbf{D} Gated Multi-Feature Fusion (GMFF) module for heterogeneous atomic feature integration.  \textbf{E} Overall architecture of ReGeoDTA.}
	\label{Fig.1}
\end{figure*}

	Identifying molecules that bind disease-relevant proteins with sufficient affinity is a central step in drug discovery~\cite{dta_1,dta_2,dta_3}. Experimental measurements of binding affinity are indispensable but remain costly and difficult to scale across the vast chemical and protein spaces that must be explored~\cite{dta_4, dta_5}. Drug--target affinity (DTA) prediction offers a computational means of estimating interaction strength and prioritizing compounds before experimental validation. Its usefulness therefore depends on whether models can capture the molecular information that is relevant to binding, rather than merely exploit recurring entities or dataset-specific associations.
	
	Machine-learning approaches to DTA prediction have progressively incorporated richer sources of molecular and protein information. Early methods, including KronRLS~\cite{KronRLS} and SimBoost~\cite{SimBoost}, relied on predefined similarities and manually engineered descriptors. Deep-learning models subsequently reduced this dependence by learning drug and target representations from sequence or raw molecular inputs, as illustrated by sequence-based DTA models such as DeepDTA~\cite{DeepDTA} and AttentionDTA~\cite{AttentionDTA}, and more recent representation-learning frameworks such as Ligand-Transformer~\cite{LigandTransformer}, DTIAM~\cite{DTIAM}, and CrossFuser~\cite{CrossFuser}. Graph- and structure-aware models further expanded the input space by incorporating molecular topology, residue connectivity, predicted protein structures, and binding-site information~\cite{GraphDTA,GraphScoreDTA,DMFFDTA,MMCLKin}. However, this expansion of input information does not ensure that affinity-relevant information remains distinguishable and usable in the learned representations.
	
	Many current methods devote substantial modelling capacity to downstream drug--target interaction modules, while paying comparatively less attention to whether affinity-relevant distinctions have already been compressed or obscured during the preceding construction of drug and protein representations.
	
	This issue is evident on the drug side. Molecular recognition depends on the coordinated contribution of diverse atomic properties~\cite{Zhang2025GNNDrugDiscovery,HiGNN}, including atom identity, valence, aromaticity, hydrogen count, and local connectivity, rather than any single chemical descriptor alone. For example, nitrogen atoms in pyridine and pyrrole share the same elemental identity but exhibit distinct hydrogen configurations, valence environments, and electronic properties, leading to different chemical behaviours. However, conventional molecular graph encoders often combine heterogeneous atomic attributes into a shared representation, leaving their individual contributions and context-dependent interactions to be recovered implicitly by subsequent neural layers~\cite{Zhang2025GNNDrugDiscovery}. Such compression may obscure chemically meaningful distinctions that characterize local molecular environments and are relevant to differential binding affinity~\cite{HimGNN}.
	
	A similar information-preservation challenge exists in protein representation. Structure-aware DTA models increasingly exploit spatial information through residue contact maps, graph connectivity or other structure-derived representations. These approaches have established the importance of protein geometry for affinity prediction; however, they often transform continuous geometric relationships into discrete structural abstractions, resulting in the loss of fine-grained spatial variations that may be relevant to molecular recognition. For example, contact-map representations~\cite{GLCNDTA,GraphScoreDTA} typically define residue interactions using a distance threshold, converting continuous distances into binary connectivity labels. Under an 8~\AA{} cutoff, residue pairs separated by 3.0~\AA{} and 7.9~\AA{} are both treated as contacts, despite representing substantially different spatial organizations, whereas pairs separated by 7.9~\AA{} and 8.1~\AA{} are assigned different labels even though their physical distances are nearly identical. Such threshold-based discretization introduces artificial discontinuities and discards geometric variations that may contribute to ligand recognition. 
	
	These drug- and protein-side observations suggest a common upstream limitation in DTA modelling: affinity prediction may be constrained by the loss of relevant information during representation construction. We hypothesized that increasingly expressive interaction modules cannot reliably reconstruct affinity-relevant distinctions once they have been obscured during representation construction. Under this hypothesis, improving the fidelity of initial drug and protein representations should enhance affinity prediction even when the downstream prediction module remains unchanged. Furthermore, because this limitation occurs before interaction modelling, representation-preserving strategies should provide benefits beyond a specific neural architecture and remain effective across different DTA backbones. 
	
	To test this hypothesis, we developed ReGeoDTA, a representation-preserving framework designed to retain affinity-relevant information before drug--target interaction modelling. ReGeoDTA focuses on improving the fidelity of drug and protein representations rather than primarily increasing the complexity of the interaction predictor. Specifically, it preserves context-dependent chemical heterogeneity in molecular representations and continuous geometric relationships in protein structures through dedicated representation strategies. These enhanced representations are subsequently integrated for affinity prediction, allowing us to examine whether improved information preservation provides a more fundamental route toward accurate DTA modelling. 
	
	Our work differs from existing DTA approaches in three main respects. First, we identify representation-stage information loss as an overlooked upstream bottleneck in DTA prediction. Second, we develop a dual-side representation-preserving framework that preserves heterogeneous chemical information and continuous protein geometry. Third, we show that representation fidelity can serve as a transferable design principle across DTA architectures with different representation and prediction mechanisms. We assess ReGeoDTA across benchmark and cold-start settings, and use randomization tests to verify that its performance reflects genuine drug--target relationships. Component ablation, controlled representation degradation, and representation--predictor decoupling examine whether the gains arise from preserving upstream information rather than from downstream predictor complexity. Cross-architecture transfer, protein attention, and drug embedding analyses assess the generality and representation-level effects of proposed representation-preserving strategies. A retrospective EGFR-held-out screen further evaluates whether these gains translate into improved prioritization of active compounds for an unseen target. The results show that ReGeoDTA achieves strong performance across benchmark and cold-start settings, while our representation-preserving strategies consistently improves diverse DTA backbones, supporting their transferability.

\section{Results}
	
	This section evaluates ReGeoDTA on three widely used DTA benchmarks under standard supervised and cold-start settings. Benchmark experiments assess its affinity prediction performance across datasets with different scales and target distributions, while cold-start experiments and an EGFR-centred retrospective screening case evaluate its generalization in more realistic drug-discovery scenarios. To examine model reliability, we further conduct randomization tests to assess whether ReGeoDTA depends on genuine drug--target correspondence rather than spurious dataset-specific patterns. In addition, component ablation, controlled representation degradation, initial representation--predictor decoupling, and cross-architecture transfer experiments are used to quantify the contributions and transferability of GMFF and PAPE (Gated Multi-Feature Fusion and Position-Aware Protein Encoder). Finally, protein-side attention-pattern analysis and drug-side feature-space analysis are performed to characterize how ReGeoDTA reshapes spatial attention organization and target-response-oriented molecular representations. For DTA prediction, model performance was evaluated using mean squared error (MSE), concordance index (CI), mean absolute error (MAE), coefficient of determination ($R^2$), and area under the precision--recall curve (AUPR). The definitions of these metrics are provided in Supplementary Section 1, and the experiment-specific data-splitting and evaluation protocols are summarized in Supplementary Section 2.
	
\subsection{ReGeoDTA improves affinity prediction}
	
	In Table~\ref{Table1}, ReGeoDTA was evaluated on the Davis, KIBA, and BindingDB datasets against methods spanning conventional machine learning, sequence-based learning, molecular graph modelling, and predictive--generative frameworks. Across the three datasets, ReGeoDTA achieved the highest CI and the lowest MSE among the compared methods and obtained the best result in 10 of the 12 dataset--metric combinations. Accordingly, its advantage was consistently observed in both pairwise affinity ranking and numerical prediction accuracy, rather than being confined to a single benchmark or metric.

\begin{table*}[p]
		\refstepcounter{table}
	\renewcommand{\arraystretch}{1.5}
		{\large\textbf{Table 1 | Performance comparison of ReGeoDTA with the state-of-the-art on the Davis, KIBA, and BindingDB datasets}} \\[1ex]
		
		\centerline{%
		\begin{tabular*}{\textwidth}{@{\extracolsep{\fill}} lllllll}
					\hline
					\textbf{Dataset} & \textbf{Model} & \textbf{CI$\uparrow$ (Std)} & \textbf{MSE$\downarrow$ (Std)} & \textbf{$R^2$\,$\uparrow$ (Std)} & \textbf{AUPR$\uparrow$ (Std)} & \textbf{Rank$\downarrow$} \\ \hline
					
					Davis & KronRLS\cite{KronRLS} & 0.871 (0.001) & 0.379 (0.001) & 0.407 (0.005) & 0.661 (0.010) & 11.50 \\ \hline
					& SimBoost\cite{SimBoost} & 0.872 (0.002) & 0.282 (0.001) & 0.644 (0.006) & 0.709 (0.008) & 9.75 \\ \hline
					& DeepDTA\cite{DeepDTA} & 0.878 (0.004) & 0.261 (0.001) & 0.630 (0.017) & 0.714 (0.010) & 9.00 \\ \hline
					& WideDTA\cite{WideDTA} & 0.886 (0.003) & 0.262 (0.009) & -- & -- & 9.50 \\ \hline
					& AttentionDTA\cite{AttentionDTA} & 0.887 (0.005) & 0.216 (0.019) & 0.677 (0.024) & \textbf{0.776 (0.024)} & 4.00 \\ \hline
					& GraphDTA\cite{GraphDTA} (GIN) & 0.893 (0.001) & 0.229 (0.001) & 0.663 (0.001) & -- & 5.00 \\ \hline
					& CoVAE\cite{CoVAE} (New-drug) & 0.712 (0.063) & 0.724 (0.096) & 0.107 (0.001) & -- & 14.33 \\ \hline
					& CoVAE\cite{CoVAE} (New-target) & 0.816 (0.011) & 0.419 (0.016) & 0.477 (0.001) & -- & 13.00 \\ \hline
					& DeepCDA\cite{DeepCDA} & 0.891 (0.003) & 0.248 (0.001) & 0.649 (0.009) & 0.739 (0.006) & 6.00 \\ \hline
					& ELECTRA-DTA\cite{ELECTRA-DTA} & 0.897 (0.003) & 0.238 (0.001) & 0.671 (0.032) & 0.698 (0.010) & 5.50 \\ \hline
					& DoubleSG-DTA\cite{DoubleSG-DTA} & 0.886 (0.003) & 0.250 (0.001) & 0.688 (0.031) & -- & 6.67 \\ \hline
					& SSM-DTA\cite{SSM-DTA} & 0.890 (0.002) & 0.219 (0.001) & -- & -- & 4.50 \\ \hline
					& GDilatedDTA\cite{GDilatedDTA} & 0.885 (0.001) & 0.237 (0.001) & 0.686 (0.001) & -- & 6.67 \\ \hline
					& DeepDTAGen\cite{DeepDTAGen} & 0.890 (0.004) & 0.214 (0.006) & 0.705 (0.001) & 0.772 (0.002) & 2.75 \\ \hline
					& \textbf{ReGeoDTA} & \textbf{0.898 (0.001)} & \textbf{0.207 (0.001)} & \textbf{0.728 (0.005)} & 0.769 (0.005) & \textbf{1.50} \\ \hline\hline
					
					KIBA & KronRLS\cite{KronRLS} & 0.782 (0.001) & 0.411 (0.001) & 0.342 (0.001) & 0.635 (0.004) & 11.75 \\ \hline
					& SimBoost\cite{SimBoost} & 0.836 (0.001) & 0.222 (0.001) & 0.629 (0.007) & 0.760 (0.003) & 10.25 \\ \hline
					& DeepDTA\cite{DeepDTA} & 0.863 (0.002) & 0.194 (0.001) & 0.630 (0.009) & 0.788 (0.004) & 9.00 \\ \hline
					& WideDTA\cite{WideDTA} & 0.875 (0.001) & 0.179 (0.008) & -- & -- & 9.50 \\ \hline
					& AttentionDTA\cite{AttentionDTA} & 0.882 (0.004) & 0.155 (0.003) & 0.755 (0.017) & 0.829 (0.005) & 4.75 \\ \hline
					& GraphDTA\cite{GraphDTA} (GIN) & 0.891 (0.001) & 0.147 (0.001) & 0.687 (0.001) & -- & 4.67 \\ \hline
					& CoVAE\cite{CoVAE} (New-drug) & 0.742 (0.011) & 0.416 (0.004) & 0.358 (0.001) & -- & 13.00 \\ \hline
					& CoVAE\cite{CoVAE} (New-target) & 0.741 (0.010) & 0.421 (0.010) & 0.356 (0.001) & -- & 14.00 \\ \hline
					& DeepCDA\cite{DeepCDA} & 0.889 (0.002) & 0.176 (0.001) & 0.682 (0.008) & 0.812 (0.005) & 6.50 \\ \hline
					& ELECTRA-DTA\cite{ELECTRA-DTA} & 0.889 (0.002) & 0.162 (0.001) & 0.727 (0.004) & 0.795 (0.006) & 5.50 \\ \hline
					& DoubleSG-DTA\cite{DoubleSG-DTA} & 0.856 (0.002) & 0.164 (0.001) & 0.721 (0.009) & -- & 8.33 \\ \hline
					& SSM-DTA\cite{SSM-DTA} & 0.895 (0.001) & 0.154 (0.001) & -- & -- & 3.50 \\ \hline
					& GDilatedDTA\cite{GDilatedDTA} & 0.876 (0.001) & 0.156 (0.001) & 0.775 (0.001) & -- & 5.33 \\ \hline
					& DeepDTAGen\cite{DeepDTAGen} & 0.897 (0.003) & 0.146 (0.008) & 0.765 (0.005) & 0.843 (0.001) & 2.25 \\ \hline
					& \textbf{ReGeoDTA} & \textbf{0.902 (0.001)} & \textbf{0.136 (0.001)} & \textbf{0.800 (0.002)} & \textbf{0.850 (0.002)} & \textbf{1.00} \\ \hline\hline
					
					BindingDB & KronRLS\cite{KronRLS} & 0.710 (0.001) & 0.910 (0.001) & -- & 0.570 (0.001) & 9.67 \\ \hline
					& DeepDTA\cite{DeepDTA} & 0.844 (0.003) & 0.633 (0.001) & 0.633 (0.004) & 0.710 (0.001) & 6.50 \\ \hline
					& AttentionDTA\cite{AttentionDTA} & 0.542 (0.001) & 0.745 (0.001) & -- & 0.698 (0.001) & 9.00 \\ \hline
					& GraphDTA\cite{GraphDTA} (GIN) & 0.858 (0.004) & 0.535 (0.008) & -- & 0.741 (0.001) & 5.33 \\ \hline
					& CoVAE\cite{CoVAE} & 0.847 (0.001) & 0.512 (0.009) & 0.412 (0.001) & 0.430 (0.001) & 6.75 \\ \hline
					& DeepCDA\cite{DeepCDA} & 0.722 (0.010) & 0.848 (0.001) & 0.531 (0.020) & 0.459 (0.030) & 8.75 \\ \hline
					& ELECTRA-DTA\cite{ELECTRA-DTA} & 0.837 (0.004) & 0.650 (0.001) & 0.670 (0.012) & -- & 7.33 \\ \hline
					& DoubleSG-DTA\cite{DoubleSG-DTA} & 0.862 (0.002) & 0.533 (0.001) & 0.726 (0.009) & -- & 4.67 \\ \hline
					& SSM-DTA\cite{SSM-DTA} & -- & 0.513 (0.001) & -- & -- & 5.00 \\ \hline
					& GDilatedDTA\cite{GDilatedDTA} & 0.868 (0.001) & 0.483 (0.002) & 0.730 (0.001) & 0.820 (0.001) & 3.00 \\ \hline
					& DeepDTAGen\cite{DeepDTAGen} & 0.876 (0.004) & 0.458 (0.002) & \textbf{0.760 (0.003)} & 0.870 (0.004) & 1.75 \\ \hline
					& \textbf{ReGeoDTA} & \textbf{0.877 (0.001)} & \textbf{0.455 (0.001)} & 0.735 (0.002) & \textbf{0.871 (0.001)} & \textbf{1.25} \\ \hline
					
		\end{tabular*}%
		}
		\vspace{0.2cm}
		
		\par\noindent {\footnotesize $\uparrow$ indicates that higher values are better, whereas $\downarrow$ indicates that lower values are better. Best performance is highlighted in bold. The (Std) is the standard deviation. Rank is the average rank across all available metrics for each model using standard competition ranking.}
		
		\label{Table1}
\end{table*}
	
	On the KIBA test set, ReGeoDTA achieved an MSE of 0.136, a CI of 0.902, and an $R^2$ of 0.800. Similarly, on the Davis test set, the model obtained an MSE of 0.207, a CI of 0.898, and an $R^2$ of 0.728. On the BindingDB test set, ReGeoDTA achieved an MSE of 0.455, a CI of 0.877, and an $R^2$ of 0.735. ReGeoDTA outperformed SimBoost, the strongest traditional machine-learning baseline, on the KIBA dataset, improving CI by 7.9\% and $R^2$ by 27.2\%, while reducing MSE by 38.7\%. Compared with the strongest deep-learning comparator for each metric, ReGeoDTA reduced MSE from 0.146 to 0.136, corresponding to a relative reduction of 6.8\%, and increased CI and $R^2$ by absolute margins of 0.005 and 0.025, respectively. It also improved AUPR from 0.843 to 0.850. A similar pattern was observed on Davis. Relative to SimBoost, ReGeoDTA improved CI and $R^2$ by 3.0\% and 13.0\%, respectively, and reduced MSE by 26.6\%. Against the strongest deep-learning comparator for each metric, ReGeoDTA reduced MSE from 0.214 to 0.207 and increased $R^2$ from 0.705 to 0.728. Its CI of 0.898 was also marginally higher than the previous best value of 0.897. However, its AUPR of 0.769 remained below the best-performing value of 0.776 achieved by AttentionDTA. The improvement on BindingDB was more modest and metric-dependent. Compared with DeepDTAGen~\cite{DeepDTAGen}, the strongest overall comparator on this dataset, ReGeoDTA improved CI from 0.876 to 0.877, reduced MSE from 0.458 to 0.455, and increased AUPR from 0.870 to 0.871. However, its $R^2$ decreased from 0.760 to 0.735. Thus, ReGeoDTA achieved its clearest and most consistent advantage on KIBA, retained improvements in affinity ranking and regression accuracy on Davis, and provided smaller but favourable aggregate gains on BindingDB.
	
	These benchmark results are consistent with the design of ReGeoDTA, which preserves drug-side chemical heterogeneity and protein-side geometric information before interaction prediction. However, standard random-split comparisons alone cannot determine whether the observed improvement arises from representation-stage information preservation or from other architectural factors. To address this limitation, we evaluated ReGeoDTA under cold-start and randomization settings to assess its generalization and reliability, followed by controlled analyses designed to separate representation quality from downstream predictor dependence.
	
\begin{table*}[!t]
	\refstepcounter{table}
	{\large\textbf{Table 2 | Cold-start performance comparison on Davis and KIBA datasets}} \\[1ex]
	
	\renewcommand{\arraystretch}{1.25}
	\begin{tabular*}{\textwidth}{@{\extracolsep{\fill}} lllccc}
		\hline
		\textbf{Dataset} & \textbf{Scenario} & \textbf{Method} & \textbf{MSE$\downarrow$ (Std)} & \textbf{CI$\uparrow$ (Std)} & \textbf{$R^2$\,$\uparrow$ (Std)} \\ \hline
		
		Davis & Cold drug & GraphDTA\cite{GraphDTA} & 0.920 (0.029) & 0.678 (0.036) & 0.160 (0.019) \\ \hline
		& & GEFA\cite{GEFA} & 0.847 (0.012) & 0.709 (0.028) & 0.182 (0.015) \\ \hline
		& & FusionDTA\cite{FusionDTA} & 0.581 (0.094) & 0.737 (0.012) & 0.187 (0.034) \\ \hline
		& & MGraphDTA\cite{MGraphDTA} & 0.563 (0.065) & 0.729 (0.022) & 0.192 (0.021) \\ \hline
		& & NHGNN\cite{NHGNN-DTA} & 0.554 (0.091) & \textbf{0.752 (0.017)} & 0.207 (0.030) \\ \hline
		& & \textbf{ReGeoDTA} & \textbf{0.546 (0.090)} & 0.744 (0.028) & \textbf{0.255 (0.030)} \\ \hline
		
		& Cold target & GraphDTA\cite{GraphDTA} & 0.510 (0.086) & 0.729 (0.012) & 0.154 (0.014) \\ \hline
		& & GEFA\cite{GEFA} & 0.433 (0.022) & 0.759 (0.009) & 0.289 (0.016) \\ \hline
		& & FusionDTA\cite{FusionDTA} & 0.364 (0.021) & 0.826 (0.011) & 0.435 (0.023) \\ \hline
		& & MGraphDTA\cite{MGraphDTA} & 0.359 (0.023) & 0.813 (0.008) & 0.425 (0.028) \\ \hline
		& & NHGNN\cite{NHGNN-DTA} & 0.344 (0.029) & 0.855 (0.016) & 0.479 (0.021) \\ \hline
		& & \textbf{ReGeoDTA} & \textbf{0.276 (0.023)} & \textbf{0.871 (0.012)} & \textbf{0.617 (0.019)} \\ \hline
		
		& All cold & GraphDTA\cite{GraphDTA} & 0.968 (0.096) & 0.579 (0.017) & 0.026 (0.016) \\ \hline
		& & GEFA\cite{GEFA} & 0.944 (0.092) & 0.610 (0.029) & 0.032 (0.022) \\ \hline
		& & FusionDTA\cite{FusionDTA} & 0.876 (0.091) & 0.645 (0.043) & 0.072 (0.048) \\ \hline
		& & MGraphDTA\cite{MGraphDTA} & 0.874 (0.090) & 0.636 (0.021) & 0.071 (0.041) \\ \hline
		& & NHGNN\cite{NHGNN-DTA} & 0.857 (0.096) & 0.665 (0.038) & 0.087 (0.051) \\ \hline
		& & \textbf{ReGeoDTA} & \textbf{0.579 (0.127)} & \textbf{0.707 (0.049)} & \textbf{0.174 (0.053)} \\ \hline \hline
		
		KIBA & Cold drug & GraphDTA\cite{GraphDTA} & 0.471 (0.047) & 0.713 (0.002) & 0.342 (0.007) \\ \hline
		& & GEFA\cite{GEFA} & 0.464 (0.032) & 0.721 (0.003) & 0.346 (0.006) \\ \hline
		& & FusionDTA\cite{FusionDTA} & 0.429 (0.031) & 0.748 (0.005) & 0.364 (0.012) \\ \hline
		& & MGraphDTA\cite{MGraphDTA} & 0.425 (0.047) & 0.746 (0.002) & 0.366 (0.016) \\ \hline
		& & NHGNN\cite{NHGNN-DTA} & 0.385 (0.029) & 0.756 (0.007) & 0.400 (0.015) \\ \hline
		& & \textbf{ReGeoDTA} & \textbf{0.375 (0.061)} & \textbf{0.769 (0.041)} & \textbf{0.480 (0.036)} \\ \hline
		
		& Cold target & GraphDTA\cite{GraphDTA} & 0.469 (0.089) & 0.610 (0.035) & 0.368 (0.057) \\ \hline
		& & GEFA\cite{GEFA} & 0.462 (0.091) & 0.636 (0.037) & 0.362 (0.052) \\ \hline
		& & FusionDTA\cite{FusionDTA} & 0.439 (0.062) & 0.685 (0.032) & 0.390 (0.067) \\ \hline
		& & MGraphDTA\cite{MGraphDTA} & 0.435 (0.055) & 0.674 (0.028) & 0.382 (0.047) \\ \hline
		& & NHGNN\cite{NHGNN-DTA} & 0.382 (0.071) & 0.732 (0.041) & 0.452 (0.054) \\ \hline
		& & \textbf{ReGeoDTA} & \textbf{0.316 (0.023)} & \textbf{0.782 (0.018)} & \textbf{0.507 (0.052)} \\ \hline
		
		& All cold & GraphDTA\cite{GraphDTA} & 0.676 (0.113) & 0.601 (0.030) & 0.149 (0.067) \\ \hline
		& & GEFA\cite{GEFA} & 0.639 (0.065) & 0.628 (0.047) & 0.152 (0.035) \\ \hline
		& & FusionDTA\cite{FusionDTA} & 0.587 (0.086) & 0.641 (0.023) & 0.193 (0.053) \\ \hline
		& & MGraphDTA\cite{MGraphDTA} & 0.590 (0.094) & 0.626 (0.028) & 0.182 (0.012) \\ \hline
		& & NHGNN\cite{NHGNN-DTA} & 0.565 (0.094) & 0.649 (0.037) & 0.218 (0.047) \\ \hline
		& & \textbf{ReGeoDTA} & \textbf{0.524 (0.036)} & \textbf{0.680 (0.016)} & \textbf{0.293 (0.041)} \\ \hline
		
	\end{tabular*}
	
	\vspace{2ex}
	
	\noindent {\footnotesize $\uparrow$ indicates that higher values are better, whereas $\downarrow$ indicates that lower values are better. Best performance is highlighted in bold. The (Std) is the standard deviation.}
	
	\label{Table_Cold_Start}
\end{table*}

\subsection{ReGeoDTA generalizes to unseen drugs and targets}
	
	To assess whether ReGeoDTA can generalize beyond observed drug--target pairs, we evaluated its performance under cold-start settings. Following the experimental protocol and baseline results reported in NHGNN-DTA~\cite{NHGNN-DTA}, we considered three scenarios: cold-drug, where test drugs are absent from the training set; cold-target, where test proteins are unseen during training; and all-cold, where neither the test drugs nor the test targets appear in the training set. Compared with random splitting, these settings provide a more stringent assessment by limiting the model's reliance on entity-level memorization.
	Table~\ref{Table_Cold_Start} compares ReGeoDTA with previous deep-learning-based DTA models under the three cold-start settings. ReGeoDTA achieved strong cold-start performance on both the Davis and KIBA datasets. On Davis, ReGeoDTA obtained the lowest MSE and the highest $R^2$ in the cold-drug setting, although NHGNN-DTA achieved a slightly higher CI. Under the cold-target and all-cold settings, ReGeoDTA outperformed all compared methods across all reported metrics. The advantages were more consistent on KIBA, where ReGeoDTA achieved the best MSE, CI and $R^2$ under all three cold-start settings. These results suggest that ReGeoDTA generalizes more effectively to unseen drugs and targets
	
\begin{figure*}[t]
	\centering
	\includegraphics[width=\textwidth]{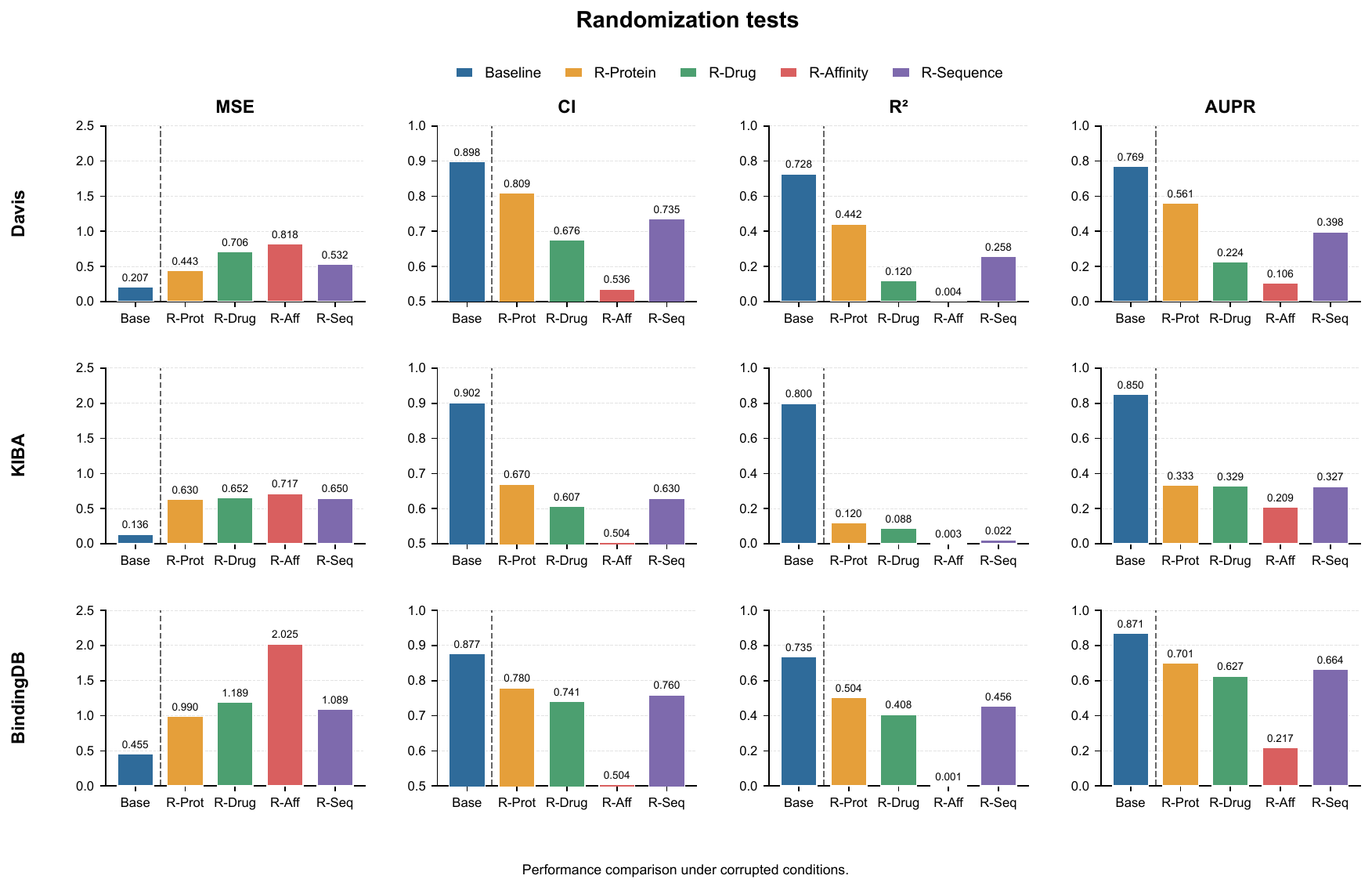}
	\caption{\textbf{Randomization tests.} Performance comparison under four corrupted conditions.}
	\label{randomization_fig}
\end{figure*}
	
\subsection{Randomization tests confirm dependence on intact drug–target relationships}
	
	To examine whether ReGeoDTA captures meaningful drug--target affinity relationships rather than relying on random correspondences or dataset-specific artifacts, we conducted four controlled randomization experiments. Within each dataset, we randomly permuted proteins, drugs, or affinity values, and additionally replaced real protein sequences with randomly generated sequences. These perturbations disrupted drug--target pairing, sample--label correspondence, or protein sequence information.
	As shown in Fig.~\ref{randomization_fig}, all randomization strategies resulted in substantial performance degradation across datasets. Permuting affinity values caused the strongest degradation: on Davis, KIBA, and BindingDB, CI decreased from 0.898, 0.902, and 0.877 in the original setting to 0.536, 0.504, and 0.504, respectively, while $R^2$ decreased to near-zero values of 0.004, 0.003, and 0.001. AUPR also dropped markedly from 0.769, 0.850, and 0.871 to 0.106, 0.209, and 0.217. Drug, protein, and sequence randomization also impaired performance, with MSE increasing consistently across all three datasets. For example, on Davis, KIBA, and BindingDB, MSE increased to 0.706, 0.652, and 1.189 under drug permutation, and to 0.532, 0.650, and 1.089 after replacing real protein sequences with randomly generated sequences. These results indicate that correct affinity labels, native drug--target pairing, and real protein sequence information all contribute to ReGeoDTA's predictive performance.
	Overall, the clear degradation under all corrupted conditions indicates that ReGeoDTA's benchmark performance cannot be attributed to random associations or spurious dataset-specific correlations. Instead, ReGeoDTA captures affinity-relevant drug--target relationships, supporting the validity of its representation-preserving design.
	
\subsection{Performance gains arise from preserving representations}
	
	We next examined whether the observed gains can be attributed to representation-stage information preservation. To this end, we conducted three complementary analyses: component ablation, controlled representation degradation, and initial representation--predictor decoupling. Together, these analyses assess whether ReGeoDTA benefits from preserving drug- and protein-side information before interaction prediction, rather than from a single architectural component or downstream prediction head. To enable controlled comparisons among these mechanism-oriented analyses while reducing dependence on a single fixed partition, we used the repeated random-split protocol described in Supplementary Section 2.
	
\begin{table*}[!t]
	\refstepcounter{table}
	{\large\textbf{Table 3 | Performance comparison of different component modifications on Davis, KIBA, and BindingDB datasets}} \\[1ex]
	
	\renewcommand{\arraystretch}{1.3}
	\begin{tabular*}{\textwidth}{@{\extracolsep{\fill}} lllllll}
		\hline
		\textbf{Dataset} & \textbf{Experiments} & \textbf{MSE$\downarrow$} & \textbf{CI$\uparrow$} & \textbf{MAE$\downarrow$} & \textbf{$R^2\uparrow$} & \textbf{AUPR$\uparrow$} \\ \hline
		Davis & \textbf{Full Model} & \textbf{0.186} & \textbf{0.906} & \textbf{0.211} & \textbf{0.756} & \textbf{0.807} \\ \hline
		& Exp. 1 & 0.187 & 0.904 & 0.211 & 0.755 & 0.806 \\ \hline
		& Exp. 2 & 0.188 & 0.901 & 0.212 & 0.751 & 0.807 \\ \hline
		& Exp. 3 & 0.200 & 0.895 & 0.222 & 0.722 & 0.782 \\ \hline
		& Exp. 4 & 0.212 & 0.895 & 0.232 & 0.689 & 0.775 \\ \hline
		& Exp. 5 & 0.195 & 0.899 & 0.218 & 0.724 & 0.786 \\ \hline
		& Exp. 6 & 0.191 & 0.904 & 0.213 & 0.728 & 0.794 \\ \hline \hline
		
		KIBA & \textbf{Full Model} & \textbf{0.127} & \textbf{0.904} & \textbf{0.188} & \textbf{0.806} & \textbf{0.847} \\ \hline
		& Exp. 1 & 0.127 & 0.903 & 0.187 & 0.806 & 0.845 \\ \hline
		& Exp. 2 & 0.130 & 0.901 & 0.190 & 0.801 & 0.844 \\ \hline
		& Exp. 3 & 0.133 & 0.902 & 0.192 & 0.781 & 0.839 \\ \hline
		& Exp. 4 & 0.138 & 0.894 & 0.199 & 0.774 & 0.832 \\ \hline
		& Exp. 5 & 0.147 & 0.893 & 0.209 & 0.743 & 0.823 \\ \hline
		& Exp. 6 & 0.135 & 0.897 & 0.196 & 0.776 & 0.836 \\ \hline \hline
		
		BindingDB & \textbf{Full Model} & \textbf{0.409} & \textbf{0.885} & \textbf{0.355} & \textbf{0.763} & \textbf{0.885} \\ \hline
		& Exp. 1 & 0.413 & 0.885 & 0.356 & 0.759 & 0.883 \\ \hline
		& Exp. 2 & 0.412 & 0.881 & 0.356 & 0.758 & 0.882 \\ \hline
		& Exp. 3 & 0.420 & 0.879 & 0.363 & 0.733 & 0.867 \\ \hline
		& Exp. 4 & 0.431 & 0.880 & 0.373 & 0.745 & 0.874 \\ \hline
		& Exp. 5 & 0.419 & 0.882 & 0.364 & 0.739 & 0.871 \\ \hline
		& Exp. 6 & 0.416 & 0.885 & 0.360 & 0.741 & 0.873 \\ \hline
	\end{tabular*}
	
	\vspace{2ex}
	
	\noindent
	\begin{minipage}{\textwidth}
		\footnotesize
		$\uparrow$ indicates that higher values are better, whereas $\downarrow$ indicates that lower values are better. Results for the full model are highlighted in bold.
		\textbf{Ablation settings:} Full Model denotes the complete model; Exp. 1 removes orientation information while retaining relative distance information; Exp. 2 removes PAPE; Exp. 3 replaces the parallel gated fusion with a cascaded fusion structure; Exp. 4 simplifies drug representations to atom-symbol embeddings; Exp. 5 replaces GMFF with conventional one-hot encoding; Exp. 6 removes the adaptive gating mechanism.
	\end{minipage}
	
	\label{Table2}
\end{table*}
	
\subsubsection{Component ablation reveals contributions of drug and protein representation modules}
	
	We first assessed the contributions of the drug- and protein-side representation modules using six model variants (Table~\ref{Table2}). For the protein encoder, we removed orientation information while retaining relative residue distances (Exp. 1), removed PAPE entirely (Exp. 2), or replaced the parallel gated fusion with cascaded fusion (Exp. 3). For the drug encoder, we reduced the input to atom-symbol embeddings (Exp. 4), replaced GMFF with conventional one-hot atom encoding (Exp. 5), or removed adaptive gating while retaining the heterogeneous atomic features (Exp. 6). All remaining model components were kept unchanged.
	
	The complete model showed the strongest overall performance across the ablation study. Removing orientation information while retaining relative residue distances produced only marginal changes across datasets, suggesting that orientation provides complementary geometric cues but contributes less substantially than relative spatial proximity. Removing PAPE entirely resulted in a small but consistent loss across the three datasets. For example, $R^2$ decreased from 0.756 to 0.751 on Davis, from 0.806 to 0.801 on KIBA, and from 0.763 to 0.758 on BindingDB. Replacing the parallel gated fusion in the protein encoder with cascaded fusion consistently reduced prediction accuracy, indicating that sequential feature integration weakened the effectiveness of multi-source protein representation learning. The drug-side variants produced larger performance changes. Simplifying the molecular representation to atom-symbol embeddings increased MSE from 0.186 to 0.212 on Davis and from 0.409 to 0.431 on BindingDB. On KIBA, replacing GMFF with one-hot encoding increased MSE from 0.127 to 0.147 and reduced $R^2$ from 0.806 to 0.743. Removing adaptive gating also impaired overall performance across the three datasets. Thus, the benefit of GMFF derives from both retaining chemically distinct atomic descriptors and integrating them adaptively, whereas PAPE provides an additional, more modest contribution from protein geometry.

	These results show that the drug- and protein-side modules make complementary contributions, but component ablation alone cannot distinguish whether the losses arise from reduced information content or from changes in network structure. We therefore progressively degraded the representations themselves.
	
\subsubsection{Cumulative representation degradation reduces overall DTA performance}
	
\begin{figure*}[!t]
	\centering
	\includegraphics[width=\textwidth]{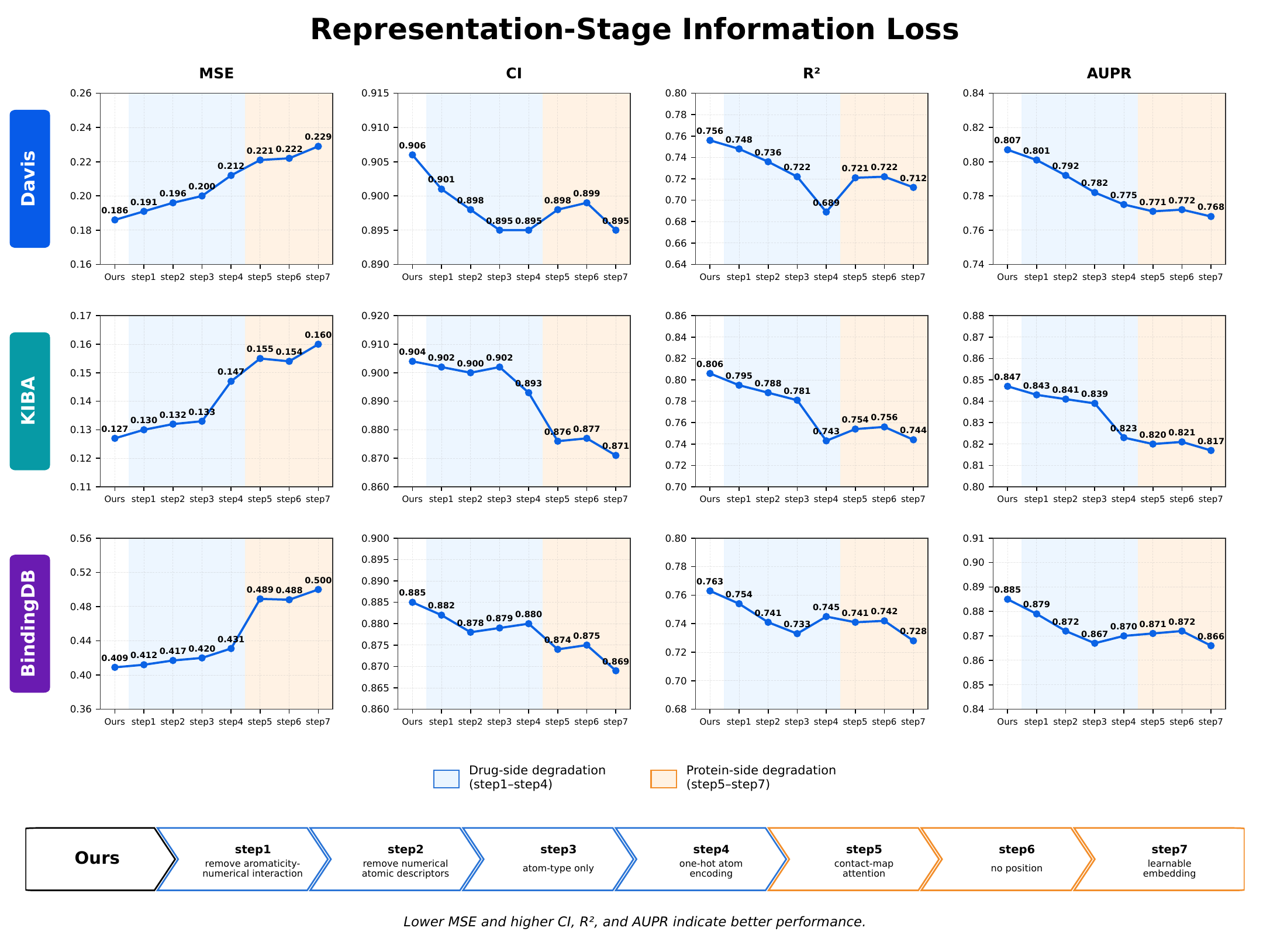}
	\caption{
			\textbf{Controlled representation degradation analysis across Davis, KIBA, and BindingDB.} 
			The complete ReGeoDTA model is used as the reference setting.
			Step 1 removes the interaction between aromaticity-related information and numerical atomic features;
			Step 2 removes numerical atomic descriptors;
			Step 3 removes aromaticity information and retains only atom-type information;
			Step 4 replaces the proposed drug encoding with conventional one-hot atom encoding;
			Step 5 replaces relative position-aware protein attention with contact-map-based attention;
			Step 6 removes protein positional information;
			and Step 7 replaces ProtTrans-derived protein representations with standard learnable embeddings.
			Lower MSE and higher CI, $R^2$, and AUPR indicate better performance.
			The overall degradation pattern suggests that preserving chemical heterogeneity, continuous protein geometry, and pretrained protein contextual information is important for affinity prediction.
		}
	\label{fig:information_loss}
\end{figure*}

	Starting from the complete ReGeoDTA model, we constructed a cumulative sequence of seven representation degradation steps (Fig.~\ref{fig:information_loss}). On the drug side, we sequentially removed aromaticity--numerical feature interactions (Step 1), numerical atomic descriptors (Step 2), and aromaticity information (Step 3), before replacing GMFF with conventional one-hot atom encoding (Step 4). On the protein side, we then replaced continuous relative geometric encoding with contact-map-based attention (Step 5), removed explicit positional information (Step 6), and finally replaced the pretrained protein representation with standard learnable embeddings (Step 7).
	Drug-side degradation produced an overall increase in prediction error. Across the first four steps, MSE increased from 0.186 to 0.212 on Davis, from 0.127 to 0.147 on KIBA, and from 0.409 to 0.431 on BindingDB. CI, $R^2$, and AUPR also generally decreased, indicating that the progressive removal of heterogeneous atomic information impaired both numerical affinity estimation and the ranking of drug--target pairs.
	Further weakening the protein representation produced an additional overall decline. At the final degradation step, MSE reached 0.229, 0.160, and 0.500 on Davis, KIBA, and BindingDB, respectively, compared with 0.186, 0.127, and 0.409 for the complete model. The corresponding $R^2$ values decreased from 0.756 to 0.712, from 0.806 to 0.744, and from 0.763 to 0.728. Thus, replacing continuous geometry and pretrained protein context with progressively coarser representations further limited affinity prediction.
	The intermediate changes were not strictly monotonic for every metric. In particular, atom-type and one-hot representations, as well as contact-map-based and position-free protein encodings, occasionally exchanged their relative ordering. These local differences may reflect changes in encoding form and optimization behaviour rather than a simple relationship between the number of input descriptors and predictive utility. For contact-map-based variants, hard discretization may also introduce restrictive or noisy structural constraints.

	Nevertheless, the consistent deterioration from the complete representation to the most degraded setting across all three datasets shows that representation-stage loss of chemical heterogeneity, protein geometry, and pretrained contextual information constrains downstream prediction. Because the degradation sequence is cumulative, it tests the consequence of joint information loss rather than assigning an independent effect size to every information source. Even so, it extends the component ablation by showing that performance generally declined as representation components were removed or coarsened before drug-target interaction modelling.
	
\subsubsection{More complex predictors do not compensate for weaker initial representations}
	
	We next examined whether a more expressive downstream predictor could compensate for weaker initial representations. The enhanced ReGeoDTA representations and GIN-based initial representations derived from GraphDTA were each paired with four prediction heads: a linear readout, cross-attention, low-rank bilinear fusion, and pairwise matching fusion (Table~\ref{Table_Representation_Predictor}). The enhanced representations outperformed the corresponding GIN-based representations across all matched comparisons. This advantage remained even when predictor complexity favoured the GIN-based setting. A linear predictor operating on the enhanced representations achieved MSE values of 0.186, 0.127, and 0.409 on Davis, KIBA, and BindingDB, respectively. These values were lower than those obtained by the strongest GIN-based predictor, pairwise matching fusion, which achieved corresponding MSE values of 0.218, 0.152, and 0.486.

\begin{table*}[!t]
		\refstepcounter{table}
		{\large\textbf{Table 4 | Initial representation--predictor decoupling analysis on Davis, KIBA, and BindingDB datasets}} \\[1ex]
		
	\renewcommand{\arraystretch}{1.3}
	\begin{tabular*}{\textwidth}{@{\extracolsep{\fill}} lllcccc}
				\hline
				\textbf{Dataset} & \textbf{Initial Representation} & \textbf{Predictor} & \textbf{MSE$\downarrow$} & \textbf{CI$\uparrow$} & \textbf{$R^2\uparrow$} & \textbf{AUPR$\uparrow$} \\ \hline
				
				Davis & Enhanced initial representation & Linear & 0.186 & 0.906 & 0.756 & 0.807 \\ \hline
				& & Cross-attention & 0.211 & 0.895 & 0.685 & 0.771 \\ \hline
				& & LRBF & 0.189 & 0.904 & 0.749 & 0.803 \\ \hline
				& & PMF & 0.184 & 0.907 & 0.761 & 0.809 \\ \hline
				& GIN-based initial representation & Linear & 0.225 & 0.898 & 0.718 & 0.770 \\ \hline
				& & Cross-attention & 0.229 & 0.896 & 0.712 & 0.767 \\ \hline
				& & LRBF & 0.221 & 0.900 & 0.724 & 0.775 \\ \hline
				& & PMF & 0.218 & 0.902 & 0.729 & 0.778 \\ \hline \hline
				
				KIBA & Enhanced initial representation & Linear & 0.127 & 0.904 & 0.806 & 0.847 \\ \hline
				& & Cross-attention & 0.146 & 0.894 & 0.744 & 0.826 \\ \hline
				& & LRBF & 0.129 & 0.903 & 0.798 & 0.844 \\ \hline
				& & PMF & 0.126 & 0.905 & 0.811 & 0.849 \\ \hline
				& GIN-based initial representation & Linear & 0.158 & 0.873 & 0.749 & 0.819 \\ \hline
				& & Cross-attention & 0.161 & 0.871 & 0.744 & 0.816 \\ \hline
				& & LRBF & 0.154 & 0.876 & 0.756 & 0.825 \\ \hline
				& & PMF & 0.152 & 0.878 & 0.760 & 0.828 \\ \hline \hline
				
				BindingDB & Enhanced initial representation & Linear & 0.409 & 0.885 & 0.763 & 0.885 \\ \hline
				& & Cross-attention & 0.434 & 0.881 & 0.743 & 0.874 \\ \hline
				& & LRBF & 0.417 & 0.883 & 0.754 & 0.880 \\ \hline
				& & PMF & 0.411 & 0.884 & 0.759 & 0.883 \\ \hline
				& GIN-based initial representation & Linear & 0.495 & 0.870 & 0.734 & 0.869 \\ \hline
				& & Cross-attention & 0.501 & 0.868 & 0.728 & 0.865 \\ \hline
				& & LRBF & 0.489 & 0.873 & 0.741 & 0.874 \\ \hline
				& & PMF & 0.486 & 0.875 & 0.745 & 0.877 \\ \hline
	\end{tabular*}
		
		\vspace{2ex}
		
		\noindent {\footnotesize $\uparrow$ indicates that higher values are better, whereas $\downarrow$ indicates that lower values are better. Enhanced initial representation denotes the strengthened drug and protein representations produced by the proposed representation modules. GIN-based initial representation denotes the representation derived from GraphDTA. LRBF denotes low-rank bilinear fusion, and PMF denotes pairwise matching fusion.}
		
		\label{Table_Representation_Predictor}
\end{table*}
	
	Once the enhanced representations were used, the linear, low-rank bilinear, and pairwise matching predictors produced broadly similar results. Pairwise matching fusion provided only small improvements on Davis and KIBA, whereas the linear predictor performed best on BindingDB. Cross-attention did not improve performance under either representation setting in this experiment. Thus, increasing interaction-model complexity was not sufficient to recover the information absent from the weaker GIN-based representations and provided limited additional benefit once the drug and protein information had already been organized effectively.
	
	Together, the ablation, degradation, and predictor-decoupling analyses support a representation-centred explanation for the performance of ReGeoDTA. Preserving heterogeneous atomic properties and continuous protein geometry improves the information available for affinity prediction, whereas downstream architectural complexity cannot fully reconstruct information that has already been lost during representation construction.
	
\subsection{Representation-preserving strategy enhances DTA prediction across model architectures}
	The preceding analyses showed that the performance gains of ReGeoDTA arise primarily from improved drug and protein representations rather than from the complexity of its downstream predictor. We next examined whether these representation-level benefits were confined to the ReGeoDTA architecture or transferable to other DTA backbones. To this end, we incorporated the GMFF-based drug encoding into six drug-side backbones---GCN, GAT, GIN, GCN+GAT, GS-DTA and DeepDTAGen---and PAPE-based protein encoding to DeepDTA, AttentionDTA and GIN. For each baseline, the original backbone and prediction head were retained, and only the representation component required to incorporate GMFF or PAPE was modified. Additional linear projection layers were introduced only when required to match feature dimensions. The original and enhanced variants were evaluated using the same data splits and training settings on Davis, KIBA and BindingDB.
	
\begin{figure*}[t]
	\centering
	\includegraphics[width=\textwidth]{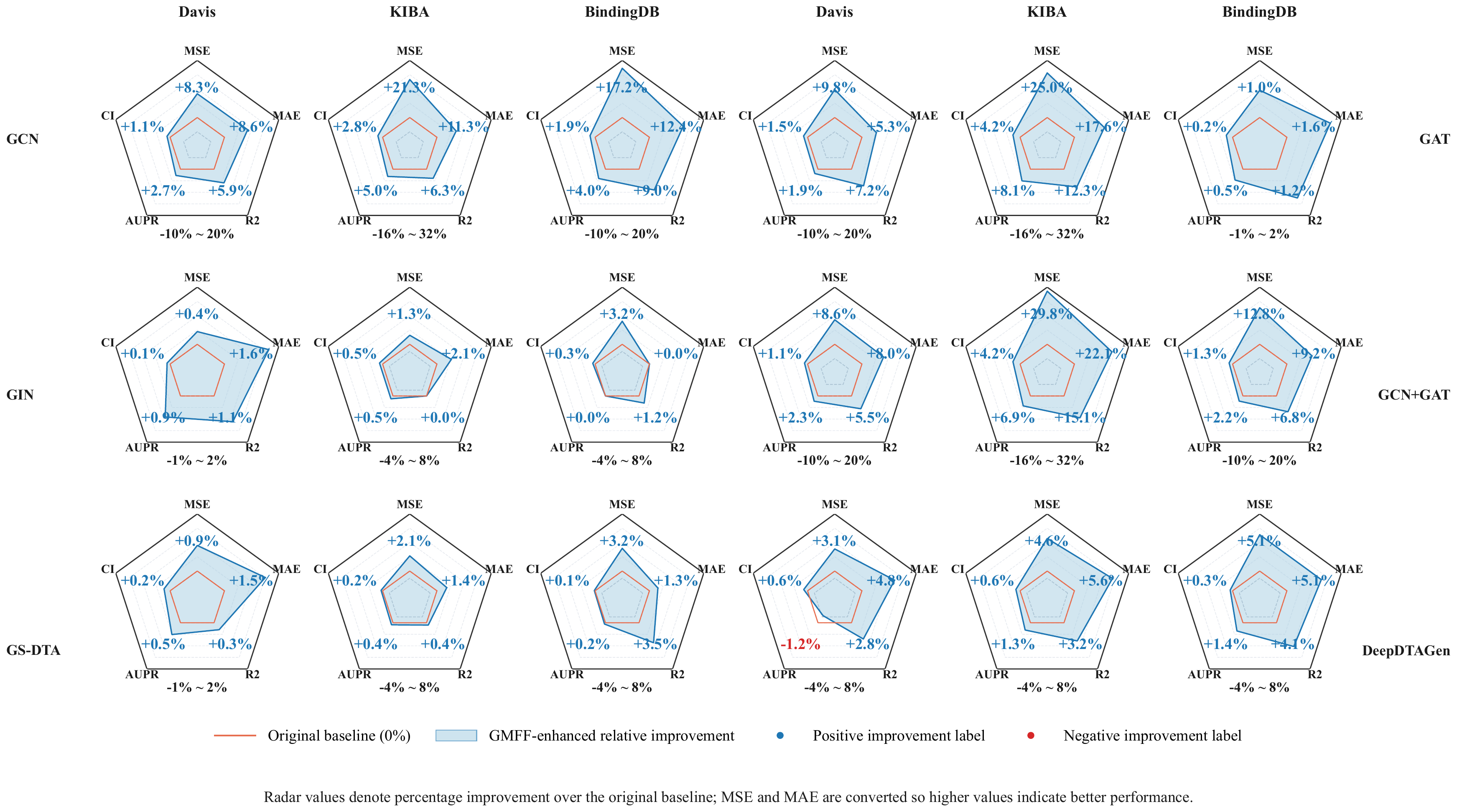}
	\caption{\textbf{Generalization of our drug encoding strategy across different graph models.} In the radar plots, values farther from the centre indicate better performance.}
	\label{drug_compare_fig}
\end{figure*}
	
\subsubsection{GMFF improves affinity prediction across diverse drug-side architectures} 
	Incorporating GMFF produced improvements across nearly all drug-encoder--dataset--metric combinations (Fig.~\ref{drug_compare_fig}). Among the 90 comparisons across six architectures, three datasets, and five evaluation metrics, 86 showed improved performance, three were essentially unchanged, and only one declined. The gains were most pronounced on KIBA. For example, in the GCN+GAT backbone, GMFF reduced MSE and MAE by 29.8\% and 22.1\%, respectively, and increased CI, $R^2$, and AUPR by 4.2\%, 15.1\%, and 6.9\%. The GAT and GCN variants showed similarly broad improvements on this dataset. On Davis, GMFF-enhanced DeepDTAGen improved MSE, MAE, CI, and $R^2$, although AUPR decreased slightly by 1.2\%. Improvements also remained evident on BindingDB, although their magnitude depended on the underlying backbone. Notably, the GMFF-enhanced variants used the same five atomic attributes across all architectures. The observed gains are therefore unlikely to arise simply from increasing descriptor dimensionality. Instead, they support the importance of retaining chemically distinct atomic information and integrating these features according to their local molecular context. Thus, GMFF did not affect all architectures equally, but its recurrent benefits across models with different graph operations indicate that heterogeneous atomic feature preservation is not restricted to the original ReGeoDTA encoder.

\subsubsection{PAPE provides consistent gains across DTA backbones}
	
\begin{figure}[!h]
	\centering
	\includegraphics[width=\columnwidth]{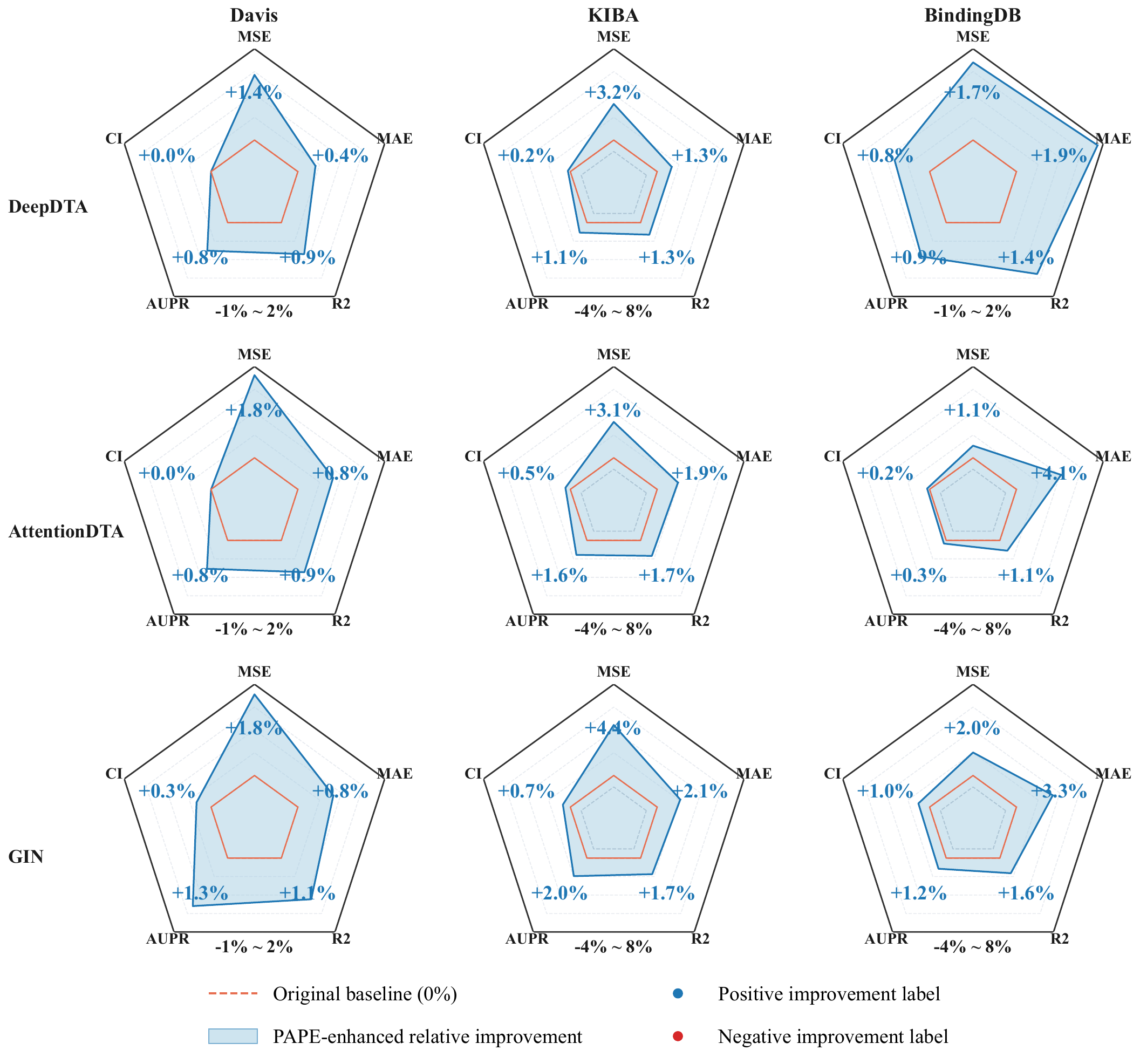}
	\caption{\textbf{Generalization of the position-aware encoder across baseline models.} In the radar plots, values farther from the centre indicate better performance.}
	\label{protein_compare_fig}
\end{figure}

	Incorporating PAPE improved performance across nearly all protein-encoder--dataset--metric combinations (Fig.~\ref{protein_compare_fig}). Among the 45 comparisons across three architectures, three datasets, and five evaluation metrics, 43 showed improved performance, two were essentially unchanged, and none declined. Although the improvements were generally smaller than those obtained with GMFF, they were more consistent across architectures and datasets. The gains were most evident on KIBA. For example, in the GIN backbone, PAPE reduced MSE by 4.4\% and increased AUPR by 2.0\%, whereas the corresponding DeepDTA and AttentionDTA variants also improved across all reported metrics. On BindingDB, PAPE improved all five metrics for each backbone, with the largest changes being a 4.1\% MAE reduction for AttentionDTA and a 3.3\% MAE reduction for GIN. On Davis, the CI values of the PAPE-enhanced DeepDTA and AttentionDTA variants remained essentially unchanged, whereas the remaining metrics improved. These results indicate that relative geometric positional encoding provides a broadly complementary signal across different protein representation schemes. Rather than acting as an architecture-specific modification, PAPE consistently enhanced models based on sequence convolution, attention, and graph representation, supporting its transferability beyond the original ReGeoDTA framework.

	\begin{table*}[!t]
	\refstepcounter{table}
	{\large\textbf{Table 5 | Performance comparison of different position encoding methods on the Davis dataset}} \\[1ex]
	\centerline{%
		\renewcommand{\arraystretch}{1.3}
		\begin{tabular*}{\textwidth}{@{\extracolsep{\fill}} lllllll}
			\hline
			\textbf{Model} & \textbf{MSE$\downarrow$} & \textbf{CI$\uparrow$} & \textbf{AUPR$\uparrow$} & \textbf{ADC} & \textbf{AS} & \textbf{AHS} \\ \hline
			Exp. 1 & 0.195 & 0.900 & 0.798 & -0.1948(0.0082) & 8.87\%(0.05\%) & 7.6 : 0.4 : 0.0 : 0.0 \\ \hline
			Exp. 2 & 0.197 & 0.898 & 0.802 & -0.1875(0.0334) & 8.70\%(0.34\%) & 6.2 : 1.0 : 0.8 : 0.0 \\ \hline
			Exp. 3 & 0.190 & 0.902 & 0.798 & -0.1449(0.1234) & 8.34\%(0.32\%) & 4.0 : 1.2 : 2.8 : 0.0 \\ \hline
			Exp. 4 & 0.195 & 0.899 & 0.797 & -0.1583(0.1128) & 8.21\%(0.09\%) & 4.0 : 3.6 : 1.4 : 0.0 \\ \hline
			Exp. 5 & 0.196 & 0.901 & 0.799 & -0.1775(0.0548) & 9.81\%(0.11\%) & 6.2 : 0.8 : 1.0 : 0.0 \\ \hline
			Exp. 6 & 0.201 & 0.897 & 0.799 & -0.1410(0.0271) & 11.57\%(0.96\%) & 3.0 : 4.6 : 0.4 : 0.0 \\ \hline
			\textbf{Exp. 7} & \textbf{0.186} & \textbf{0.906} & \textbf{0.807} & \textbf{-0.3919(0.0072)} & \textbf{10.74\%(0.07\%)} & \textbf{8.0 : 0.0 : 0.0 : 0.0} \\ \hline
		\end{tabular*}%
	}
	\vspace{2ex}
	\par\noindent {\footnotesize $\uparrow$ indicates that higher values are better, whereas $\downarrow$ indicates that lower values are better. Standard deviations are shown in parentheses.}
	\label{Table3}
\end{table*}	

	Together, the cross-architecture experiments reveal complementary transfer patterns: GMFF produced larger gains by reorganizing heterogeneous atomic information, whereas PAPE produced smaller but more consistent improvements by introducing relative protein geometry. The recurrence of these benefits across DTA models with different representation and prediction mechanisms supports representation preservation as a general upstream design principle rather than an architecture-specific advantage of ReGeoDTA.

\subsection{Relative geometric encoding aligns protein attention with residue-level geometry}
	To clarify why PAPE improves affinity prediction, we tested whether its benefit comes from adding positional information alone or from aligning attention with spatially proximal residues. On the Davis dataset, we compared seven otherwise matched protein encoders: an attention encoder without positional encoding (Exp. 1), sinusoidal absolute encoding (Exp. 2), three-dimensional coordinate-based encoding (Exp. 3), coordinate concatenation (Exp. 4), radial basis function-based relative encoding (Exp. 5), PPE-based absolute encoding (Exp. 6)~\cite{PPE}, and an encoder using the relative geometric positional encoding of PAPE (Exp. 7). Predictive performance was assessed using MSE, CI, and AUPR.
	
	We further characterized the resulting attention patterns using three complementary measures. Attention--distance correlation (ADC) measures the Spearman correlation between attention weights and pairwise residue distances, with more negative values indicating a stronger association between high attention and spatial proximity. Attention sparsity (AS) quantifies the proportion of attention mass assigned to the ten most strongly attended key positions for each query residue. Attention-head specialization (AHS) summarizes how the eight attention heads are distributed across focused, weakly focused, neutral, and long-range spatial-preference categories. Detailed definitions, equations, and calculation procedures for these measures are provided in Supplementary Section 4.
	
	Relative geometric positional encoding produced the best predictive performance among the seven strategies (Table~\ref{Table3}). PAPE (Exp. 7) achieved the lowest MSE of 0.186 and the highest CI and AUPR of 0.906 and 0.807, respectively. By comparison, the position-free baseline (Exp. 1) obtained an MSE of 0.195, a CI of 0.900, and an AUPR of 0.798. The remaining positional strategies (Exp. 2--6) yielded MSE values of 0.190--0.201, CI values of 0.897--0.902, and AUPR values of 0.797--0.802. Thus, introducing positional or structural information did not consistently improve prediction. This pattern indicates that the predictive benefit depended not merely on the presence of positional information, but on how geometric relationships were represented and incorporated into the attention mechanism.

	The predictive advantage of PAPE (Exp. 7) was accompanied by the strongest attention--distance alignment. PAPE produced an ADC of -0.3919 $\pm$ 0.0072, compared with -0.1948 $\pm$ 0.0082 for the position-free baseline (Exp. 1) and values ranging from -0.1875 to -0.1410 for the alternative positional strategies (Exp. 2--6). The magnitude of this negative correlation was therefore approximately twice that of the position-free model, indicating that higher attention weights were more consistently associated with spatially proximal residue pairs. PAPE also produced an AHS profile of 8.0 : 0.0 : 0.0 : 0.0, placing all eight attention heads in the focused category. Although the position-free encoder was also dominated by focused heads, with a profile of 7.6 : 0.4 : 0.0 : 0.0, its weaker ADC and lower predictive performance indicate that local attention can emerge without explicit geometric encoding but was less strongly aligned with three-dimensional residue proximity.
	
	Attention concentration alone did not account for the predictive advantage of PAPE. PPE-based absolute encoding (Exp. 6) produced the highest AS value of 11.57 $\pm$ 0.96\%, exceeding the 10.74 $\pm$ 0.07\% obtained with PAPE (Exp. 7), but yielded the highest MSE and the lowest CI among the tested strategies. Thus, concentrating attention on a small number of positions was not sufficient to improve affinity prediction. Rather, the advantage of PAPE was associated with stronger geometric alignment of attention rather than greater attention concentration alone.
	
	Together, these results associate the predictive advantage of PAPE with a more geometrically aligned organization of protein attention. Because ADC, AS, and AHS characterize spatial organization rather than ligand-binding specificity, we therefore examined whether the resulting attention patterns corresponded to structurally plausible residue relationships in experimentally resolved protein--ligand complexes.
	
\begin{figure*}[!t]
	\centering
	\includegraphics[width=\textwidth]{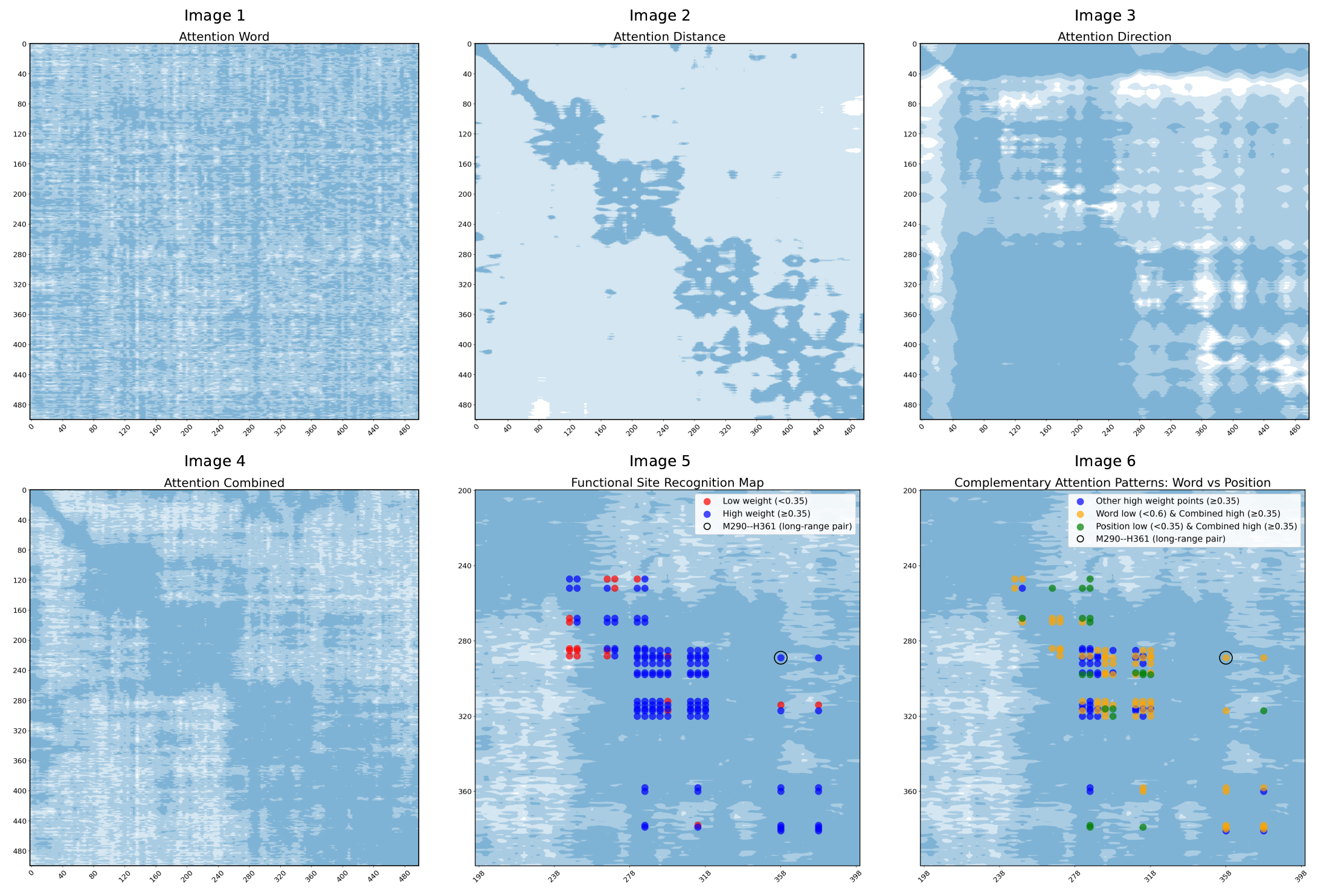}
	\caption{\textbf{Decomposition and structural localization of PAPE-derived attention in the ABL1--Nilotinib complex (PDB ID: 3CS9).} Images 1--4 show the semantic, distance-aware, orientation-aware, and fused residue-pair attention maps, respectively, illustrating how the attention pattern changes as geometric information is incorporated. Image 5 evaluates the fused map against 22 annotated binding-site residues assigned to four structurally defined regions, with blue indicating high fused attention ($\geq 0.35$) and red indicating lower fused attention ($<0.35$). Image 6 analyses component complementarity: blue denotes other high-weight fused pairs, orange denotes high-weight pairs identified through positional complementarity, and green denotes high-weight pairs identified through semantic complementarity. The black open circle marks the representative M290--H361 long-range cross-region association.}
	\label{case_study_fig}
\end{figure*}

\begin{figure*}[!t]
	\centering
	\includegraphics[width=\textwidth]{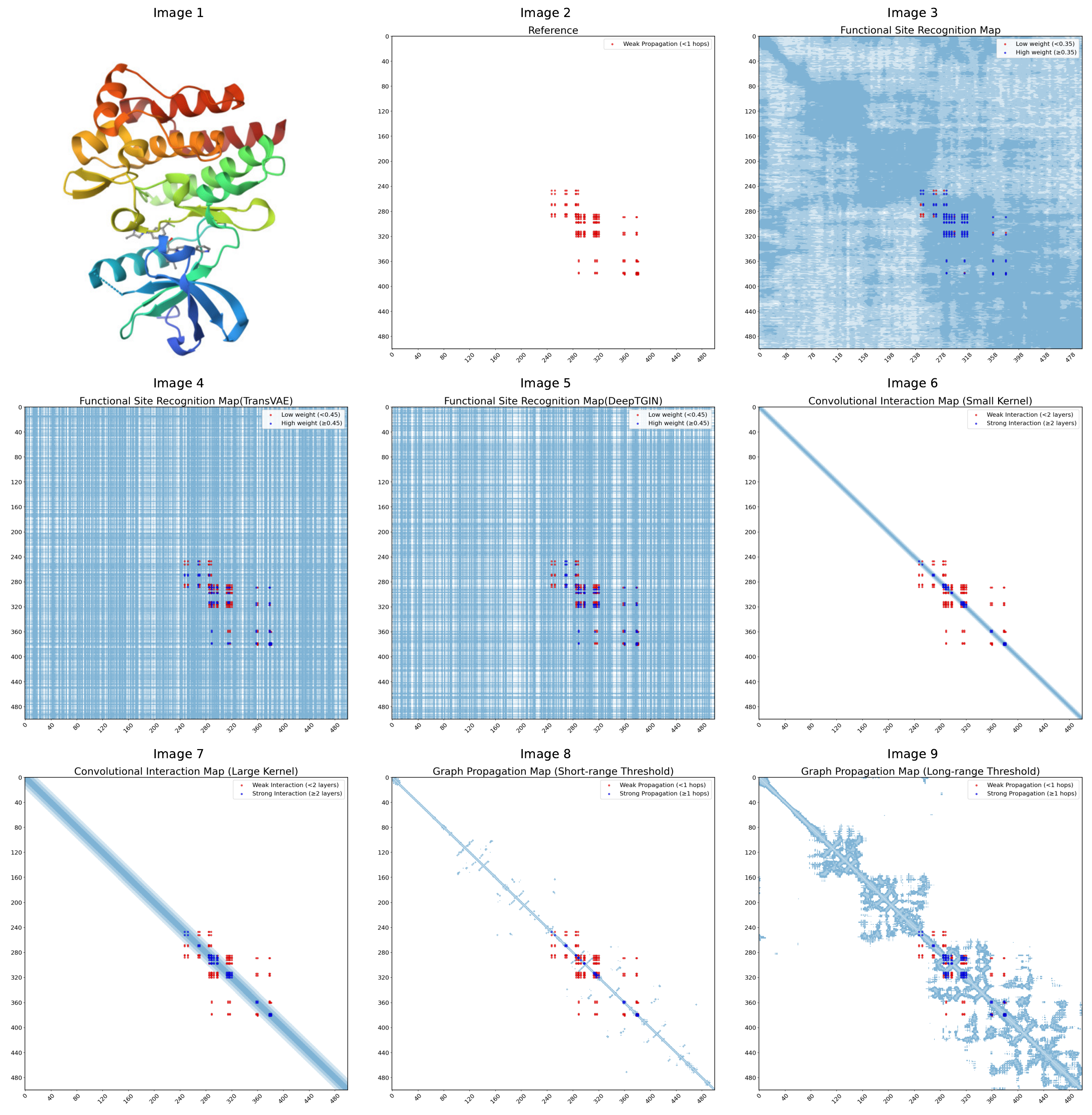}
	\caption{\textbf{Attention, convolutional and graph-propagation maps represent different internal quantities and are presented for qualitative structural comparison (PDB ID: 3CS9).} \textbf{Image 1:} Overall structure of ABL1. \textbf{Image 2:} Binding site with spatial functional groups. \textbf{Image 3:} Attention map from our proposed method. \textbf{Image 4:} Attention map from TransVAE. \textbf{Image 5:} Attention map from DeepTGIN. \textbf{Image 6:} Interaction map from CNN with a small convolution kernel. \textbf{Image 7:} Interaction map from CNN with a large convolution kernel. \textbf{Image 8:} Graph neural network propagation map under a low-threshold protein contact map. \textbf{Image 9:} Propagation map under a high-threshold contact map.}
	\label{case_study_fig2}
\end{figure*}

\begin{figure*}[!t]
	\centering
	\includegraphics[width=\textwidth]{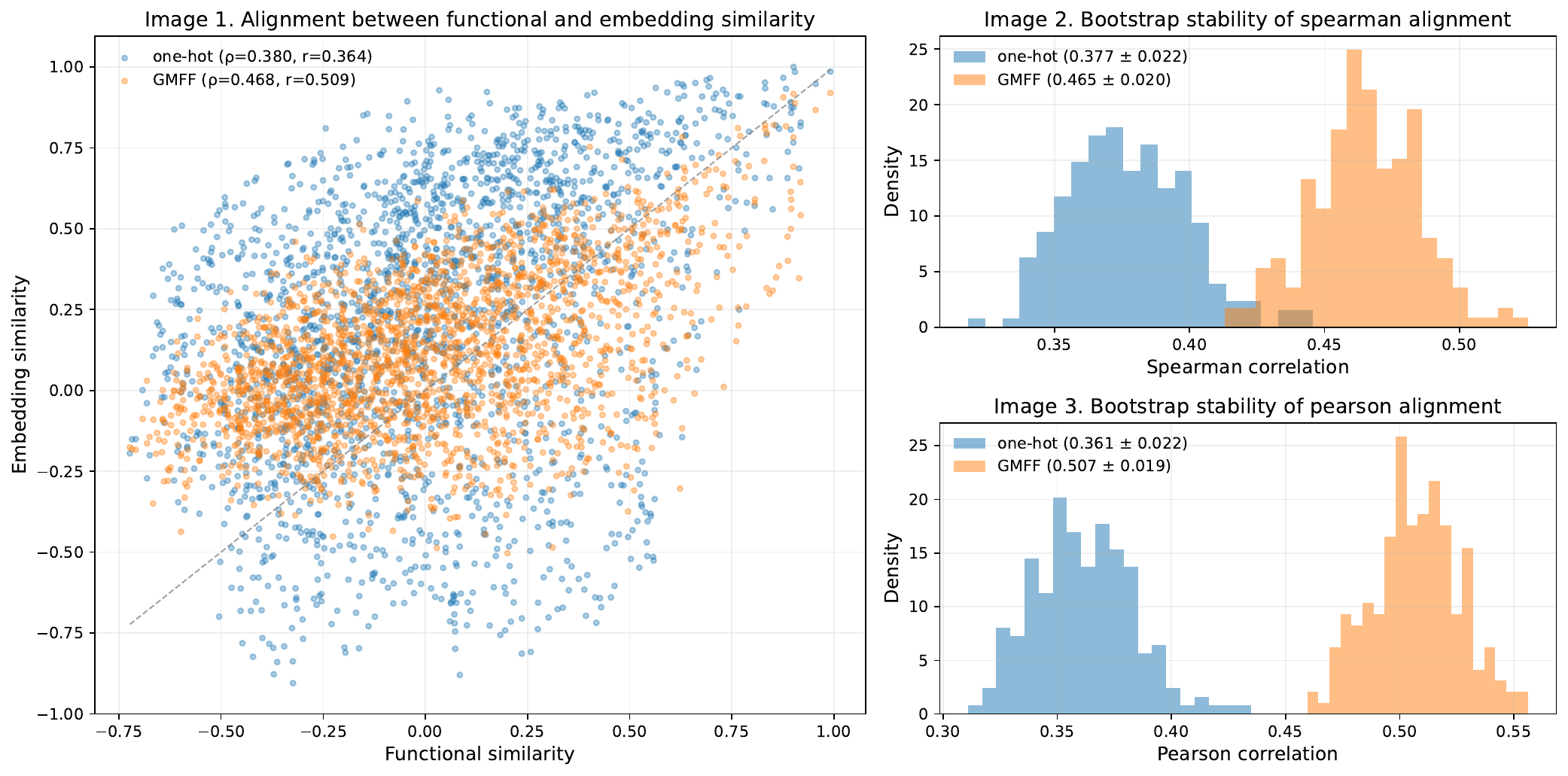}
	\caption{Alignment between drug embedding similarity and multi-target affinity-profile similarity on the Davis dataset. \textbf{Image 1:} Alignment between functional and embedding similarity.  \textbf{Image 2:} Bootstrap stability of spearman alignment. \textbf{Image 3:} Bootstrap stability of pearson alignment. }
	\label{fig:drug_embedding_comparison}
\end{figure*}

\begin{table*}[t]
	\refstepcounter{table}
	{\large\textbf{Table 6 | Retrospective target-held-out prioritization of EGFR candidates under a top-10\% screening budget}} \\[1ex]
	\centerline{%
		\renewcommand{\arraystretch}{1.3}
		\begin{tabular*}{\textwidth}{@{\extracolsep{\fill}} lcccccccccc}
			\hline
			\textbf{Model}
			& \textbf{MSE$\downarrow$}
			& \textbf{MAE$\downarrow$}
			& \textbf{CI$\uparrow$}
			& \textbf{Spearman$\uparrow$}
			& \textbf{Pearson$\uparrow$}
			& \textbf{Prec.$\uparrow$}
			& \textbf{Rec.$\uparrow$}
			& \textbf{F1$\uparrow$}
			& \textbf{Top-10\% Hits$\uparrow$}
			& \textbf{Top-10\% EF$\uparrow$} \\ \hline
			
			DeepDTAGen 
			& 1.421 
			& 0.976 
			& 0.735 
			& 0.643 
			& 0.604 
			& \textbf{0.636} 
			& 0.452 
			& 0.528 
			& 6/11 
			& 1.918 \\ \hline
			
			ReGeoDTA 
			& \textbf{1.244} 
			& \textbf{0.826} 
			& \textbf{0.748} 
			& \textbf{0.683} 
			& \textbf{0.708} 
			& 0.488 
			& \textbf{0.677} 
			& \textbf{0.568} 
			& \textbf{9/11} 
			& \textbf{2.877} \\ \hline
		\end{tabular*}
	}
	\vspace{2ex}
	\par\noindent {\footnotesize $\uparrow$ indicates that higher values are better, whereas $\downarrow$ indicates that lower values are better. The EGFR screening set contained 109 compounds, of which 31 met the experimental high-activity threshold of pIC50 $\geq$ 7.5. Predicted high-activity candidates were defined using the same threshold on predicted pIC50. Prec., Rec., and F1 denote the precision, recall, and F1 score for identifying high-activity candidates under this threshold. Top-10\% corresponds to the top 11 compounds among the 109 EGFR candidates, and Top-10\% Hits denotes the number of high-activity compounds among these top-ranked candidates. EF denotes the enrichment fold relative to random selection.}
	\label{Table_EGFR_screening}
\end{table*}

\subsection{PAPE-derived attention aligns with known binding-site organization in a representative protein--ligand complex}
	To determine whether this geometry-aligned attention also corresponded to known binding-site organization, we examined the ABL1--Nilotinib complex (PDB ID: 3CS9), for which experimentally resolved structural information and binding-site annotations are available. Twenty-two annotated binding-site residues were grouped into four structurally defined regions: the hinge/P-loop, the gatekeeper/hydrophobic-spine core, the catalytic-loop/hydrophobic-spine extension, and the activation-loop-bottom region. In Fig.~\ref{case_study_fig}, we first decomposed PAPE-derived attention into semantic, distance-aware, orientation-aware, and fused components (Images 1--4), and then examined how the fused pattern corresponded to the annotated binding-site organization (Images 5 and 6). Images 1--4 should be interpreted comparatively: broadly distributed patterns indicate less selective emphasis across residue pairs, whereas localized high-weight regions indicate stronger prioritization of particular residue relationships. Image 5 evaluates whether residue pairs formed by the annotated binding-site residues receive high fused attention, while Image 6 further identifies high-weight pairs whose prominence emerges from the complementarity between semantic and position-related information. The detailed procedures for residue selection, attention normalization, binding-region coverage analysis, and component-complementarity analysis in the attention visualization experiment are provided in Supplementary Section 6.
	
	The individual attention components captured different but incomplete aspects of residue organization. Semantic attention (Image 1) produced broadly distributed associations driven by residue-feature similarity, without explicitly encoding three-dimensional proximity. Distance-aware attention (Image 2) produced more spatially organized patterns around geometrically proximal residue pairs, but proximity alone did not distinguish binding-related relationships from generic structural neighbourhoods. Orientation-aware attention (Image 3) further introduced directional dependence, although its high-weight regions remained distributed across multiple structural contexts. After these signals were fused (Image 4), the attention pattern became more selective, with reduced diffuse background associations and several localized high-weight residue-pair regions remaining prominent. Image 5 further relates this fused pattern to the annotated binding-site residues by showing which pairwise positions among these residues fall into the high-attention category. Image 6 then shows that some high-weight pairs become prominent when semantic and position-related information are considered jointly, suggesting that relative geometry refines, rather than merely amplifies, the content-based attention pattern.
	
	The fused attention map also retained cross-region associations that could not be explained by sequence proximity alone. A representative example is the M290--H361 pair (black open circle in Fig. 6, Images 5 and 6): M290 belongs to the gatekeeper/hydrophobic-spine core, whereas H361 belongs to the catalytic-loop/hydrophobic-spine extension. Despite their large separation along the protein sequence, the highlighted M290--H361 relationship received high fused attention. This observation does not by itself establish a direct physical interaction between M290 and H361. Rather, it shows that PAPE can assign high importance to long-range residue relationships that are compatible with the three-dimensional organization of the annotated binding region.
	
	We next examined whether similarly selective residue-pair patterns emerged from other protein-encoding paradigms (Fig.~\ref{case_study_fig2}). Image 1 provides the overall ABL1 structure, Image 2 provides the reference map of annotated binding-site residue relationships, and Image 3 shows the PAPE-derived attention map. TransVAE and DeepTGIN were included as self-attention-based models (Images 4 and 5); three-layer CNNs with kernel sizes of 4 and 12 represented sequence-local encoders with small and large receptive fields, respectively (Images 6 and 7); and three-layer GNNs constructed using 5~\AA{} and 10~\AA{} contact thresholds represented predefined structure-based propagation models (Images 8 and 9). All models were applied to the same ABL1--Nilotinib complex, and their residue-pair maps were visualized using a common residue ordering. Because attention weights, convolutional interaction maps, and graph-propagation maps represent different internal quantities, the comparison concerns the organization and selectivity of the residue-pair patterns rather than their numerical magnitudes.

	The different encoder families produced visually distinct residue-pair patterns. The self-attention maps produced by TransVAE and DeepTGIN (Images 4 and 5) contained structured but broadly distributed row- and column-wise patterns. These patterns indicate that individual residues can receive attention across many partners, but provide limited selectivity for particular residue pairs around the annotated binding-site reference. The CNN-derived maps (Images 6 and 7) were instead dominated by diagonal bands because their receptive fields constrain information exchange primarily to nearby sequence positions. Increasing the kernel size expanded these bands along the sequence axis, enlarging the accessible sequence neighbourhood without selectively concentrating the pattern around the annotated binding-site relationships. The contact-map GNNs (Images 8 and 9) incorporated three-dimensional proximity, but their propagation patterns largely followed the connectivity imposed by the selected distance threshold. Increasing the threshold from 5~\AA{} to 10~\AA{} introduced more structurally permitted edges and consequently expanded the propagation pattern across the residue-pair map.
	
	In comparison, the PAPE-derived map (Image 3) was not dominated by broadly distributed row- and column-wise structures, sequence-local diagonal bands, or globally expanded threshold-defined propagation patterns. Instead, its high-weight residue pairs showed a more selective concentration around the annotated binding-site relationships while retaining both local and cross-region associations. This difference reflects how structural information is used. CNN kernel sizes define the sequence range over which information can be exchanged, and GNN contact thresholds define the structural edges over which information can propagate. PAPE instead incorporates relative distance and orientation into learned attention, allowing the importance of each residue pair to depend jointly on its semantic and geometric context. Accordingly, this comparison provides qualitative evidence of different residue-pair organization rather than a quantitative ranking of model interpretability.
	
	Together, this case study connects the global attention--distance alignment observed in Table 5 with a concrete structural example. PAPE did not merely increase attention to spatially proximal residues; it combined residue-feature information with relative geometry to produce a more selective pattern around known binding-site regions. These results provide case-level structural support for the position-aware protein representation learned by ReGeoDTA.

\subsection{GMFF-derived drug embeddings better reflect multi-target affinity profiles}
	
	We next examined whether GMFF's predictive gains were accompanied by a more affinity-relevant organization of the drug representation space. Specifically, we tested whether drugs with similar affinity patterns across a common target panel were also represented more similarly in the learned embedding space. We used the Davis dataset because its relatively dense drug--target affinity matrix provides each drug with a comparable affinity profile across the same set of protein targets. For each drug pair, we calculated the cosine similarity between their affinity-profile vectors and between their learned embedding vectors, and quantified the correspondence between the two similarity measures using Spearman and Pearson correlations. GMFF-derived embeddings were compared with embeddings obtained from an otherwise matched one-hot drug representation. Bootstrap resampling was used to assess the stability of the observed correspondence. Detailed procedures for affinity-profile preprocessing, embedding extraction, similarity calculation, correlation analysis, and bootstrap resampling are provided in Supplementary Section 7.
	
	As shown in Fig.~\ref{fig:drug_embedding_comparison}, GMFF-derived embeddings showed a stronger correspondence with multi-target affinity similarity than one-hot embeddings (Image 1). The Spearman correlation between affinity-profile similarity and embedding similarity increased from 0.3801 with one-hot encoding to 0.4677 with GMFF, whereas the corresponding Pearson correlation increased from 0.3636 to 0.5093. Thus, drug pairs with more similar affinity patterns across the Davis target panel were more consistently represented as similar under GMFF.
	This pattern remained stable under bootstrap resampling. The mean Spearman correlation increased from 0.3771 $\pm$ 0.0221 for one-hot embeddings to 0.4652 $\pm$ 0.0198 for GMFF-derived embeddings, producing a clear rightward shift in the bootstrap distribution (Image 2). The corresponding Pearson estimates increased from 0.3613 $\pm$ 0.0218 to 0.5071 $\pm$ 0.0190, showing the same stability pattern in the Pearson-based analysis (Image 3).
	Together, these results indicate that GMFF organizes drug embeddings in a manner that more closely reflects multi-target affinity behaviour than conventional one-hot encoding. This representation-level correspondence is consistent with the interpretation that preserving chemically distinct atomic attributes and integrating them adaptively retains affinity-relevant distinctions that are weakened by conventional atom-type one-hot encoding. The resulting organization of the drug representation space provides a possible explanation for the predictive and cross-architecture gains observed with GMFF.

\subsection{ReGeoDTA improves real discovery}
	
	To determine whether the predictive gains of ReGeoDTA translate into more effective candidate prioritization for an unseen target, we performed a retrospective target-held-out screening analysis using EGFR as the held-out target. All EGFR-associated interactions were excluded from model training, and the 109 EGFR-associated compounds were retained as an independent held-out screening set. DeepDTAGen was evaluated under the same protocol as a strong reference model. Compounds with experimentally measured pIC50 $\geq$ 7.5 were defined as high-activity candidates, resulting in 31 active compounds, and all candidates were ranked according to their predicted pIC50 values. We evaluated global affinity prediction using regression and ranking metrics, threshold-based identification using precision, recall, and F1 score, and early retrieval under a top-10\% screening budget corresponding to 11 candidates. Detailed procedures for EGFR held-out data construction, activity-threshold selection, and enrichment-factor calculation are provided in Supplementary Section 8.
	
	ReGeoDTA showed stronger overall agreement with the experimentally measured EGFR affinities (Table~\ref{Table_EGFR_screening}). Compared with DeepDTAGen, it reduced MSE from 1.421 to 1.244 and MAE from 0.976 to 0.826, while increasing CI from 0.735 to 0.748, Spearman correlation from 0.643 to 0.683 and Pearson correlation from 0.604 to 0.708. These improvements indicate that ReGeoDTA more accurately estimated affinity magnitudes and preserved the relative ordering of EGFR candidates.
	
	At the predefined pIC50 threshold, ReGeoDTA identified 21 of the 31 high-activity compounds, compared with 14 identified by DeepDTAGen. This increased recall from 0.452 to 0.677 and F1 score from 0.528 to 0.568. The gain in active-compound recovery was accompanied by a reduction in precision from 0.636 to 0.488, indicating that ReGeoDTA assigned high-activity predictions to a broader set of compounds. Thus, threshold-based classification revealed a recall--precision trade-off: ReGeoDTA missed fewer experimentally active compounds, at the cost of including more inactive compounds at the fixed activity threshold.
	
	The ranking-based analysis showed a clearer advantage under the constrained screening budget. Among the top 11 candidates, ReGeoDTA retrieved nine high-activity compounds, corresponding to a hit rate of 81.8\%, whereas DeepDTAGen retrieved six, corresponding to a hit rate of 54.5\%. The resulting enrichment factor increased from 1.918 to 2.877, indicating that the ReGeoDTA-ranked subset contained high-activity compounds at nearly 2.9 times the rate expected under random selection. This stronger early enrichment is particularly relevant when only a small fraction of the candidate library can be advanced to experimental validation.
	
	Together, these results provide retrospective evidence that the improvements in affinity estimation and ranking achieved by ReGeoDTA can translate into more efficient early prioritization of active compounds for an unseen target.
	
\section{Discussion}
	
	Existing DTA studies have often improved affinity prediction by developing more expressive drug--target interaction modules. Complementing this line of work, this study examined whether information loss during drug and protein representation construction can also limit drug--target affinity prediction. ReGeoDTA was developed to preserve two sources of information that are commonly simplified before interaction modelling: heterogeneous atomic properties on the drug side and continuous relative geometry on the protein side. Across benchmark, cold-start, representation-degradation, and predictor-decoupling analyses, the results support a representation-centred explanation for its performance. Progressive removal of atomic heterogeneity, relative protein geometry, and pretrained protein context generally impaired prediction, whereas replacing a simple prediction head with more expressive interaction modules did not compensate for weaker initial representations. Conversely, once the enhanced representations were available, several prediction heads produced broadly similar performance. These findings suggest that downstream interaction modelling is not a substitute for sound upstream representation construction: interaction modules can reorganize available information, but cannot reliably reconstruct affinity-relevant distinctions that have already been compressed or discarded. Preserving affinity-relevant chemical and geometric information before interaction modelling may therefore provide an additional direction for improving DTA prediction.
	
	The drug- and protein-side analyses further support this representation-centred view: GMFF produced larger gains by preserving chemically distinct atomic attributes and organizing drug embeddings according to multi-target affinity behaviour, whereas PAPE produced smaller but more consistent improvements by introducing relative protein geometry into attention. The successful transfer of GMFF and PAPE to graph-based drug encoders and sequence-, attention-, and graph-based protein encoders further indicates that these benefits are not specific to the native ReGeoDTA architecture, but reflect representation-preserving strategies that can complement alternative DTA backbones.
	
	The randomization and EGFR-held-out analyses further tested whether ReGeoDTA's performance gains reflected meaningful drug--target relationships and retrospective candidate prioritization. Disrupting drug--target pairing, affinity labels, or protein sequence information markedly impaired performance, indicating that ReGeoDTA did not rely on random associations alone. In the EGFR-held-out screening case, these gains translated into stronger global affinity estimation and improved early enrichment of experimentally high-activity compounds under a top-10\% screening budget.
	
	Together, these findings suggest that preserving chemical and geometric information before interaction prediction is a useful complement to increasingly expressive DTA architectures.
	
	Several limitations define the scope of the present conclusions. First, PAPE relies on predicted protein structures, so inaccuracies in local geometry, flexible regions, or alternative conformational states may propagate into the relative geometric positional encoding. Incorporating structural confidence or conformational ensembles may help address this limitation. Second, GMFF uses a selected set of atomic attributes. Richer physicochemical or quantum-chemical descriptors may provide additional information, but they may also introduce redundancy, dataset-specific bias, and greater computational cost. Finally, the EGFR experiment was retrospective and target-held-out, rather than a prospective screen of a fully external compound library. It therefore demonstrates improved prioritization within the available labelled data, but not experimental discovery of previously uncharacterized binders.

\section{Methods}
\subsection{Datasets}
	
	We evaluated our model on three benchmark datasets: Davis~\cite{Davis}, KIBA~\cite{KIBA}, and BindingDB~\cite{BindingDB}, which are widely used benchmarks for drug--target affinity prediction.
	
	The Davis dataset contains 68 drugs and 442 targets, with 30,056 labelled drug--target pairs. The original affinity measurements are reported as $K_d$ values and were converted to p$K_d$ values as the prediction labels, ranging from 5.0 to 10.8:
\begin{equation}
		pK_d = -\log_{10}\left(\frac{K_d}{1 \times 10^9}\right)
\end{equation}
	
	The KIBA dataset contains 2,116 drugs and 229 targets, with 118,254 labelled drug--target pairs represented by KIBA scores, ranging from 0.0 to 17.2.
	
	The BindingDB dataset contains 18,044 drugs and 1,620 targets, with 56,525 labelled drug--target pairs. The original affinity measurements are reported as IC$_{50}$ values and were converted to pIC$_{50}$ values as the prediction labels, providing a larger and more diverse benchmark for evaluation.
	
\subsection{Experimental setup}
	
	For the benchmark comparison experiments on Davis, KIBA, and BindingDB, we followed the data splitting strategy adopted by DeepDTA, GraphDTA, and related baseline methods to ensure direct comparability with previously reported results. Specifically, each dataset was divided into six folds, with one fold held out as the test set and the remaining five folds used for five-fold cross-validation to obtain the final model. The same splitting strategy was also applied in the randomization experiments. For the cold-start experiments, we followed the data partitions and evaluation protocol reported in NHGNN-DTA~\cite{NHGNN-DTA}.
	
	For the remaining training-based analyses, including component ablation, representation degradation, initial representation--predictor decoupling, positional-encoding comparison, and cross-architecture transfer experiments, we adopted a repeated random-split protocol to reduce dependence on a single fixed partition.
	
	\textbf{Data splitting.} Each dataset was randomly split into training and test sets with a 4:1 ratio. This process was repeated five times using five different random seeds, and the average performance across the five splits was reported as the final result. Hyperparameters were empirically set before final evaluation, and the held-out test sets were used only for reporting final performance.
	
	\textbf{Optimization strategy.} During the first half of training epochs, we maintained a fixed initial learning rate. During the second half, we applied a cosine annealing scheduler, and the model checkpoint from the final epoch was used for evaluation. This schedule was used to keep the learning rate stable during early training and gradually reduce it during later optimization.
	
	\textbf{Hyperparameter setting.} We used learning rates selected from $\{0.001, 0.0005, 0.0001\}$ and batch sizes selected from $\{64, 256\}$ according to the empirical training behaviour of each model family. To keep the number of parameter-update steps comparable across batch sizes, the number of epochs for batch size 256 was set to four times that used for batch size 64. This setup allowed different architectures to be evaluated under a comparable number of optimization updates while permitting learning rate and batch size to vary across model variants, particularly in the cross-architecture transfer experiments where GMFF or PAPE was integrated into baseline models. Experiment-specific data-splitting and evaluation protocols are summarized in Supplementary Section 2, and implementation and hyperparameter settings are provided in Supplementary Section 3.
	
\subsection{Overall architecture}
	
	As illustrated in Fig.~\ref{Fig.1}, ReGeoDTA consists of five functional blocks: a Position-Aware Protein Encoder (PAPE) with relative geometric positional encoding; a drug graph encoder with parallel GCN and GAT--GCN branches; a Gated Multi-Feature Fusion (GMFF) module for heterogeneous atomic attribute integration; a protein positional information processing module for generating processed relative geometric information from 3D protein coordinates; and an overall affinity prediction architecture that integrates the drug and protein representations for drug--target affinity estimation.
	
	On the drug side, we first convert drug SMILES strings into molecular graph representations. For each node in the drug graph, GMFF construct node representations from heterogeneous atomic attributes by combining categorical and numerical atom-level information. The GMFF module uses aromaticity-conditioned numerical augmentation and gated fusion to support adaptive integration of feature components with different scales and embedding forms. The resulting GMFF-derived drug node features are then processed by two parallel graph-encoding branches, namely a GCN branch and a GAT--GCN branch, and their outputs are concatenated to form the final drug representation.
	
	On the protein side, we generate initial residue-level embeddings using the pretrained protein language model ProtTrans and obtain predicted protein structures using ESMFold. The extracted $C\alpha$ coordinates are processed to construct relative distance and orientation matrices, which provide the continuous relative geometry used by PAPE. The ProtTrans-derived embeddings are fed into two parallel branches: a CNN branch and PAPE. Within PAPE, the key (K) and value (V) sequences are generated from down-sampled protein features, whereas the query (Q) sequence is derived from the original full-length features to retain full-length query resolution. During attention computation, relative geometric positional encoding combines content-based residue similarity with distance- and orientation-aware geometric biases. The output of PAPE is subsequently combined with the CNN output through gated fusion, followed by max pooling to obtain the final position-aware protein representation.
	
	The final drug and protein representations are concatenated and passed into a feedforward prediction module to produce the affinity score. The prediction loss is defined as the mean squared error between the predicted and ground-truth affinity values.
	
\subsection{Feature representation}
\subsubsection{Target protein encoding}
	
	Following DeepDTA~\cite{DeepDTA}, we set the maximum protein sequence length to 1200 for Davis and 1000 for KIBA and BindingDB. For sequences longer than the dataset-specific maximum length, the input protein sequence was first truncated. ProtTrans embeddings were then extracted from the original or truncated sequence, and the resulting residue embeddings were zero-padded to the dataset-specific maximum length when necessary. To obtain initial residue-level protein embeddings, we used the pretrained protein language model ProtTrans~\cite{prottrans}. We then applied a 1D convolution with kernel size 1 to project the ProtTrans embeddings from 1024 to 128 dimensions. The extracted features are represented as follows:
\begin{equation}
		\mathbf{p} = \text{ProtTrans}(P) \in \mathbb{R}^{L \times 1024}
		\label{eq:protTrans}
\end{equation}
\begin{equation}
		\mathbf{h} = \text{Conv1D}_{k=1}(\mathbf{p}) \in \mathbb{R}^{L \times 128}
\end{equation}
	where $L$ denotes the padded or truncated sequence length.
	
\subsubsection{Protein 3D structure acquisition}
	
	We obtained predicted protein structures using ESMFold~\cite{esm}. For sequences longer than the dataset-specific maximum length, structure prediction was performed on the truncated sequence. The extracted coordinate representations were then zero-padded to the dataset-specific maximum length, consistent with the processing of ProtTrans embeddings. For each residue, we extracted the three-dimensional $C\alpha$ coordinates as its structural coordinate representation. To reduce the number of geometric anchors and computational cost while retaining structural context, we applied average pooling with a kernel size of 8 and stride of 4 to generate down-sampled coordinate anchors, reducing the coordinate length to one quarter of the input length. Based on the original coordinates and the down-sampled coordinate anchors, we constructed relative distance and orientation matrices to encode the continuous relative geometry used by PAPE for relative geometric positional encoding.
	To reduce the number of geometric anchors and computational cost
	
\begin{equation}
		\mathbf{C} = \text{ESMFold}(P) \in \mathbb{R}^{L \times 3}
\end{equation}
\begin{equation}
		\mathbf{C}_{\text{down}} = \text{AvgPool}_{8,4}(\mathbf{C}) \in \mathbb{R}^{L' \times 3}
\end{equation}
\begin{equation}
		\mathbf{D} \in \mathbb{R}^{L \times L'}, \quad \mathbf{D}_{ij} = \|\mathbf{c}_i - \mathbf{c}_{\text{down},j}\|_2
\end{equation}
\begin{equation}
		\mathbf{O} \in \mathbb{R}^{L \times L' \times 3}, \quad 
		\mathbf{O}_{ij} =
		\frac{\mathbf{c}_i - \mathbf{c}_{\text{down},j}}
		{\|\mathbf{c}_i - \mathbf{c}_{\text{down},j}\|_2 + \epsilon}
\end{equation}
	where $L$ denotes the padded or truncated sequence length, $L' = L/4$ denotes the down-sampled coordinate length, $\mathbf{D}$ is the relative distance matrix, $\mathbf{O}$ is the relative orientation matrix, and $\epsilon$ is a small constant used for numerical stability.
	
\subsubsection{Gated Multi-Feature Fusion (GMFF) for drug molecular features}
	
	We take drug SMILES as input and construct molecular graphs for each drug using RDKit, following similar practices in GraphDTA~\cite{GraphDTA} and related methods. The node features include atom symbol, degree, number of hydrogen atoms, implicit valence, and aromaticity. Unlike conventional atom-type one-hot encoding, Gated Multi-Feature Fusion (GMFF) adopts a differentiated strategy for constructing heterogeneous atomic attributes. Specifically, the five atom-level features are categorized into numerical attributes, including degree, number of hydrogen atoms, and implicit valence, and categorical attributes, including atom symbol and aromaticity.
	
	Atom symbols were mapped to 128-dimensional embeddings, and the binary aromaticity indicator was mapped to a 32-dimensional embedding. For the three numerical attributes, each feature channel was independently min--max normalized using statistics computed from the training set.
	
	To model context-dependent combinations of aromaticity and numerical atomic attributes, GMFF introduces a 6-dimensional aromaticity-conditioned augmentation vector. This augmentation allows the same numerical attribute to be represented differently depending on whether the corresponding atom is aromatic or non-aromatic. This design reflects the fact that atoms with similar degree, valence, or hydrogen counts can play different chemical roles under different aromatic contexts, as illustrated by pyridinic and pyrrolic nitrogen~\cite{aromatic_nitrogen}.
	
	Specifically, if a node is aromatic, the three numerical attributes are placed in the first three dimensions of the augmentation vector, with the remaining three dimensions set to zero. Conversely, if a node is non-aromatic, the numerical attributes are placed in the last three dimensions, with the first three dimensions set to zero. The 6-dimensional augmentation vector is then concatenated with the original 3-dimensional numerical attribute vector, forming a 9-dimensional input to the feedforward network. This input is projected to a 96-dimensional numerical representation, which is subsequently concatenated with the atom-symbol and aromaticity embeddings to form a 256-dimensional vector. Finally, GMFF applies a gating mechanism to modulate the contribution of feature components with different scales and embedding forms.
	
\begin{equation}
		\mathbf{f}_{\text{num}} =
		[f_{\text{deg}}, f_{\text{hyd}}, f_{\text{val}}] \in \mathbb{R}^{3}
\end{equation}
	
\begin{equation}
		\mathbf{f}_{\text{aug}} =
		\begin{cases}
			[\mathbf{f}_{\text{num}}, \mathbf{0}_3] \in \mathbb{R}^{6}, & \text{if aromatic} \\
			[\mathbf{0}_3, \mathbf{f}_{\text{num}}] \in \mathbb{R}^{6}, & \text{otherwise}
		\end{cases}
\end{equation}
	
\begin{equation}
		\mathbf{h}_{\text{num}} =
		\text{FFN}([\mathbf{f}_{\text{num}}, \mathbf{f}_{\text{aug}}])
		\in \mathbb{R}^{96}
\end{equation}
	
\begin{equation}
		\mathbf{h} =
		[\mathbf{h}_{\text{atom}}, \mathbf{h}_{\text{aro}}, \mathbf{h}_{\text{num}}]
		\in \mathbb{R}^{256}
\end{equation}
	
\begin{equation}
		\mathbf{z} = \mathbf{h} + \mathbf{g} \odot \mathbf{h}
\end{equation}
	
	where $f_{\text{deg}}$, $f_{\text{hyd}}$, and $f_{\text{val}}$ denote the normalized degree, number of hydrogen atoms, and implicit valence, respectively. $\mathbf{h}_{\text{atom}}$ denotes the atom-symbol embedding, $\mathbf{h}_{\text{aro}}$ denotes the aromaticity embedding, $\mathbf{g}$ denotes the gating vector, and $\mathbf{z}$ is the final GMFF-derived atom representation.
	
\subsection{Network models}
	
\subsubsection{Position-Aware Protein Encoder (PAPE)}
	
	We adopted the same CNN layers as DeepDTA~\cite{DeepDTA} to learn representations from the initial protein embeddings. To incorporate continuous relative geometry into protein representation, these initial embeddings, together with the relative distance and orientation matrices, are fed into PAPE, which integrates self-attention with relative geometric positional encoding~\cite{attention}. To reduce computational cost, we applied a down-sampling strategy similar to that used in protein positional information extraction. Specifically, we employed a 1D convolution with a kernel size of 8 and stride of 4, reducing the sequence length to $L'=L/4$.
	
	The query (Q) sequence is generated from the original full-length features, while the key (K) and value (V) sequences are derived from the down-sampled features. The similarity matrix is then computed accordingly. Because the same down-sampling scale is used for both the key/value sequence and the coordinate anchors, the resulting similarity matrix and relative geometric matrices maintain a one-to-one correspondence. The residue mask is down-sampled consistently with the key/value sequence; a down-sampled position is masked if all coordinates aggregated into that position correspond to padded residues. The resulting mask is applied before the softmax operation to prevent padded positions from contributing to the attention weights.
	
\begin{equation}
		\mathbf{X}_{\text{down}} = \text{Conv1D}_{k=8, s=4}(\mathbf{X}) \in \mathbb{R}^{L' \times d}
\end{equation}
\begin{equation}
		\mathbf{Q} = \mathbf{X} \mathbf{W}^Q, \quad
		\mathbf{K} = \mathbf{X}_{\text{down}} \mathbf{W}^K, \quad
		\mathbf{V} = \mathbf{X}_{\text{down}} \mathbf{W}^V
\end{equation}
	where $\mathbf{X} \in \mathbb{R}^{L \times d}$ denotes the original sequence features, and $\mathbf{X}_{\text{down}} \in \mathbb{R}^{L' \times d}$ denotes the down-sampled sequence features. After linear projection, $\mathbf{Q}$, $\mathbf{K}$, and $\mathbf{V}$ are reshaped into $h$ attention heads.
	
	For the relative distance matrix, we applied a bucketing strategy for discretization and combined it with a trainable bias table to obtain the corresponding distance bias matrix~\cite{relative_position}. For the relative orientation matrix, we employed a feedforward network with ReLU activation to derive the orientation bias matrix.
	
	Prior to the softmax operation, the content-based similarity matrix is augmented by adding the distance bias matrix and the orientation bias matrix. The updated similarity matrix therefore combines content-based residue similarity with distance- and orientation-aware geometric biases. To assess the contribution of orientation information, we also implemented a distance-only variant by setting $\alpha=0$.
	
\begin{equation}
		\mathbf{B}_{\text{dist}} = \mathbf{E}_{\text{dist}}[\text{Bucket}(\mathbf{D})] \in \mathbb{R}^{L \times L' \times h}
\end{equation}
\begin{equation}
		\mathbf{B}_{\text{dir}} = \text{FFN}_{\text{ReLU}}(\mathbf{O}) \in \mathbb{R}^{L \times L' \times h}
\end{equation}
\begin{equation}
		\mathbf{S}^{(m)} =
		\frac{\mathbf{Q}^{(m)}{\mathbf{K}^{(m)}}^\top}{\sqrt{d_k}}
		+ \mathbf{B}_{\text{dist}}^{(m)}
		+ \alpha \cdot \mathbf{B}_{\text{dir}}^{(m)}, \quad m=1,\ldots,h
\end{equation}
\begin{equation}
		\text{Attention}^{(m)}(\mathbf{Q}, \mathbf{K}, \mathbf{V}) =
		\text{softmax}(\mathbf{S}^{(m)} + \mathbf{M})\mathbf{V}^{(m)}
\end{equation}
	where $\mathbf{E}_{\text{dist}} \in \mathbb{R}^{B \times h}$ is a learnable distance bias table with $B$ buckets, $h$ is the number of attention heads, $\text{Bucket}(\mathbf{D})$ maps each distance value to its corresponding bucket index, $\mathbf{E}_{\text{dist}}[\cdot]$ denotes the lookup operation, $\alpha$ is a learnable parameter, and $\mathbf{M}$ denotes the attention mask used to exclude padded positions.
	
	Finally, the output of PAPE is combined with the CNN output via a gated fusion mechanism, followed by max pooling to obtain the final position-aware protein representation. The CNN branch uses padding to preserve the sequence length, ensuring that its output is aligned with the PAPE output.
	
\begin{equation}
		\mathbf{H}_{\text{CNN}} = \text{CNN}(\mathbf{X}) \in \mathbb{R}^{L \times d}
\end{equation}
	
\begin{equation}
		\mathbf{H}_{\text{PAPE}} = \text{PAPE}(\mathbf{X}, \mathbf{D}, \mathbf{O}) \in \mathbb{R}^{L \times d}
\end{equation}
	
\begin{equation}
		\mathbf{H} = \mathbf{H}_{\text{CNN}} + \mathbf{g} \odot \mathbf{H}_{\text{PAPE}}
\end{equation}
	
\begin{equation}
		\mathbf{z}_{\text{prot}} = \text{MaxPool}(\mathbf{H}) \in \mathbb{R}^{d}
\end{equation}
	where $\mathbf{g} \in \mathbb{R}^{L \times d}$ is a learnable gating weight used to modulate the PAPE branch, $\odot$ denotes element-wise multiplication, and $\mathbf{z}_{\text{prot}}$ is the final position-aware protein representation.
	
\subsubsection{Drug graph network}
	
	The GMFF-derived drug node features are subsequently processed by two parallel graph-encoding branches: a GCN branch and a GAT--GCN branch.
	
	The GCN branch consists of three GCN layers. A GCN layer can be formulated as:
\begin{equation}
		\mathbf{X}_{\text{GCN}}^{(l+1)} =
		\sigma\left(
		\mathbf{D}_{\text{deg}}^{-\frac{1}{2}}
		\tilde{\mathbf{A}}
		\mathbf{D}_{\text{deg}}^{-\frac{1}{2}}
		\mathbf{X}_{\text{GCN}}^{(l)}
		\mathbf{W}_{\text{GCN}}^{(l)}
		\right)
\end{equation}
	where $\tilde{\mathbf{A}}$ is the adjacency matrix with self-connections, $\mathbf{D}_{\text{deg}}$ is the corresponding degree matrix, $\mathbf{W}_{\text{GCN}}^{(l)}$ is a trainable weight matrix, and $\sigma$ denotes the ReLU activation function.
	
	The GAT--GCN branch comprises one GAT layer followed by one GCN layer. The GAT layer is formulated as:
\begin{equation}
		\mathbf{X}_{\text{GAT},i}^{(l+1)} =
		\sigma\left(
		\sum_{j \in \mathcal{N}(i)}
		\alpha_{ij}^{(l)}
		\mathbf{W}_{\text{GAT}}^{(l)}
		\mathbf{X}_{\text{GAT},j}^{(l)}
		\right)
\end{equation}
\begin{equation}
		e_{ij}^{(l)} =
		\text{LeakyReLU}\left(
		\mathbf{a}^{(l)T}
		[
		\mathbf{W}_{\text{GAT}}^{(l)}
		\mathbf{X}_{\text{GAT},i}^{(l)}
		\, \| \,
		\mathbf{W}_{\text{GAT}}^{(l)}
		\mathbf{X}_{\text{GAT},j}^{(l)}
		]
		\right)
\end{equation}
\begin{equation}
		\alpha_{ij}^{(l)} =
		\frac{\exp(e_{ij}^{(l)})}
		{\sum_{k \in \mathcal{N}(i)} \exp(e_{ik}^{(l)})}
\end{equation}
	where $\mathcal{N}(i)$ denotes the neighbours of node $i$, including the node itself; $\mathbf{W}_{\text{GAT}}^{(l)}$ is a trainable weight matrix; $\mathbf{a}^{(l)}$ is a trainable attention parameter vector; and $\|$ denotes vector concatenation.
	
	Finally, average pooling and max pooling are applied to the node representations produced by each branch, and the pooled outputs are concatenated to form the final drug representation $\mathbf{z}_{\text{drug}}$.
	
\subsection{Positional encoding variants}
	
	To compare PAPE with alternative protein positional encoding strategies, we implemented a position-free baseline and five positional encoding variants: sinusoidal absolute position encoding, 3D coordinate-based absolute position encoding, 3D coordinate-based feature concatenation, RBF-based relative position encoding, and PPE-based absolute position encoding. The PAPE formulation is described above. These variants were compared in the experiments section to assess whether the benefit of PAPE comes from introducing positional information in general or from its relative geometric positional encoding.
	
\subsubsection{Sinusoidal absolute position encoding}
	
	Sinusoidal positional encoding~\cite{attention} is an absolute position encoding scheme implemented using sine and cosine functions. It was originally introduced in the Transformer architecture~\cite{attention}. The positional encoding $\mathbf{PE}$ is formulated as:
\begin{equation}
		\mathbf{PE}_{(\mathrm{pos}, 2i)} =
		\sin\left(
		\frac{\mathrm{pos}}{10000^{2i / d_{\text{model}}}}
		\right)
\end{equation}
\begin{equation}
		\mathbf{PE}_{(\mathrm{pos}, 2i+1)} =
		\cos\left(
		\frac{\mathrm{pos}}{10000^{2i / d_{\text{model}}}}
		\right)
\end{equation}
	where $\mathrm{pos}$ denotes the position in the sequence, $d_{\text{model}}$ is the hidden dimension, and $i$ is the dimension index.
	
\subsubsection{3D coordinate-based absolute position encoding}
	
	For 3D coordinate-based absolute position encoding, the $C\alpha$ coordinates of each residue are projected through a feedforward network into a $d$-dimensional vector and added to the residue features.
	
\subsubsection{3D coordinate-based feature concatenation}
	
	For 3D coordinate-based feature concatenation, the $C\alpha$ coordinates of each residue are projected to a 32-dimensional vector through a feedforward network and then concatenated with the residue features before being passed to the encoder. The coordinate-derived feature dimension is kept smaller than the original residue-feature dimension.
	
\subsubsection{RBF-based relative position encoding}
	
	Radial Basis Function (RBF) encoding~\cite{RBF} maps continuous distance values into a high-dimensional feature space. It can be formulated as:
\begin{equation}
		\text{RBF}_{ij,k}
		=
		\exp\left(
		-\frac{(D_{ij}-c_k)^2}{2w^2}
		\right)
\end{equation}
	where $D_{ij}$ denotes the distance between residue $i$ and down-sampled coordinate anchor $j$, consistent with the distance matrix used in PAPE, and $c_k$ is the $k$-th centre point. By uniformly sampling the distance space, we construct a set of centre points $\{c_1, c_2, \ldots, c_N\}$, which enables encoding the original distance into an $N$-dimensional feature vector. Each dimension corresponds to the response intensity for a specific centre point. When the input distance is close to a given centre point, the response value approaches 1; when the distance is far from the centre point, the response value approaches 0. The width parameter $w$ controls the sensitive region of the response function.
	
\subsubsection{PPE-based absolute position encoding}
	
	Proposed positional embedding (PPE)~\cite{PPE} was originally introduced in medical image processing, where absolute position encoding is constructed based on the geometric relationship between image pixels and the coordinate origin. It can be formulated as:
\begin{equation}
		\mathbf{PPE}_i =
		\frac{\sin\left(
			\omega \sqrt{x^2 + y^2 + z^2}
			\right)}{S}
\end{equation}
	where $x$, $y$, and $z$ denote the coordinates, $S$ is a scaling factor, and $\omega$ is the angular frequency.
	
	Because protein function is more closely associated with internal spatial organization than with an arbitrary global coordinate origin, we define the coordinate origin as the centroid of the protein structure when generating PPE. This variant is still treated as an absolute position encoding strategy because each residue is assigned an individual coordinate-origin-based positional value rather than a pairwise relative positional bias:
\begin{equation}
		\mathbf{PPE}_i =
		\frac{\sin\left(
			\omega d_i
			\right)}{S}
\end{equation}
\begin{equation}
		d_i =
		\left\|
		\mathbf{r}_i -
		\frac{1}{L}
		\sum_{j=1}^{L}
		\mathbf{r}_j
		\right\|_2
\end{equation}
	where $\mathbf{r}_i$ denotes the original coordinates of residue $i$, and $d_i$ denotes the distance from residue $i$ to the protein coordinate centroid.

\newpage
\vfill

\end{document}